\documentclass[10pt,twocolumn,twoside]{IEEEtran} 
\ifCLASSINFOpdf
\else
\fi

\usepackage{graphicx}
\usepackage{epstopdf}
\usepackage[skip=10pt, font=footnotesize]{caption}
\usepackage[skip=10pt, font=footnotesize]{subcaption}
\usepackage{amsmath}
\usepackage{amssymb}
\usepackage{enumitem}
\usepackage{cite}
\usepackage{color}
\usepackage{float}

\newlist{Properties}{enumerate}{2}
\setlist[Properties]{label=P\arabic*),itemindent=*}

\newlist{Algorithm}{enumerate}{2}
\setlist[Algorithm]{label={\bf{step}} \arabic*:,itemindent=*}

\newcommand{\tightoverset}[2]{%
  \mathop{#2}\limits^{\vbox to -.5ex{\kern-0.75ex\hbox{$#1$}\vss}}}

\IEEEoverridecommandlockouts
\begin{document}

\title{Spatio-Temporal Waveform Design for Multi-user Massive MIMO Downlink with 1-bit Receivers}

\author{Ahmet Gokceoglu, \textit{Member, IEEE}, Emil Bj\"{o}rnson, \textit{Member, IEEE}, Erik G. Larsson, \textit{Fellow, IEEE},  and \\Mikko Valkama, \textit{Senior Member, IEEE} \thanks{This work is accepted by IEEE Journals of Selected Topics in Signal Processing. Copyright (c) 2016 IEEE. Personal use of this material is permitted. However, permission to use this material for any other purposes must be obtained from the IEEE by sending a request to pubs-permissions@ieee.org.
		
This work was supported by the Academy of Finland under the projects 284694 and 288670, the Finnish Funding Agency for Technology and Innovation (Tekes) under the TAKE-5 project, the EU FP7 under ICT-619086 (MAMMOET) and ELLIIT.

A. Gokceoglu, and M. Valkama are with the Department of
Electronics and Communications Engineering, Tampere University of Technology, FI-33101 Tampere, Finland (e-mails: ahmet.gokceoglu@tut.fi and mikko.e.valkama@tut.fi). 

E.  Bj\"{o}rnson and E. G. Larsson are with the Department of Electrical Engineering (ISY), Link\"{o}ping University, Link\"{o}ping, Sweden (e-mails: emil.bjornson@liu.se and erik.g.larsson@liu.se).

Preliminary work addressing a limited subset of initial results was presented at IEEE ICC 2016  \cite{gokceoglu_waveform_2016}.
}}

\markboth{IEEE Journal of Selected Topics in Signal Processing, Accepted for Publication, 2016}%
{Spatio-Temporal Waveform Design for Multi-user Massive MIMO Downlink with 1-bit Receivers}

\maketitle

\vspace{-10pt}

%
\begin{abstract}
Internet-of-Things (IoT) refers to a high-density network of low-cost low-bitrate terminals and sensors where also low energy consumption is one central feature. As the power-budget of classical receiver chains is dominated by the high-resolution analog-to-digital converters (ADCs), there is a growing interest towards deploying receiver architectures with reduced-bit or even 1-bit ADCs.
In this paper, we study waveform design, optimization and detection aspects of multi-user massive MIMO downlink where user terminals adopt very simple 1-bit ADCs with oversampling. In order to achieve spectral efficiency higher than 1 bit/s/Hz per real-dimension, {\color{black}and per receiver antenna}, we propose a two-stage precoding structure, namely a novel quantization precoder followed by maximum-ratio transmission (MRT) or zero-forcing (ZF) type spatial channel precoder which jointly form the multi-user-multiantenna transmit waveform. The quantization precoder outputs are designed {\color{black}and optimized, under appropriate transmitter and receiver filter bandwidth constraints,} to provide controlled inter-symbol-interference (ISI) enabling the input symbols to be uniquely detected from 1-bit quantized observations with a low-complexity symbol detector in the absence of noise. {\color{black}An additional optimization constraint is also imposed in the quantization precoder design to increase the robustness against noise and residual inter-user-interference (IUI).} The purpose of the spatial channel precoder, in turn, is to suppress the IUI and provide high beamforming gains such that good symbol-error rates (SERs) can be achieved in the presence of noise and interference. 
Extensive numerical evaluations illustrate that the proposed spatio-temporal precoder based multiantenna waveform design can facilitate good multi-user link performance, despite the extremely simple 1-bit ADCs in the receivers, hence being one possible enabling technology for the future low-complexity IoT networks.
\end{abstract}


\IEEEpeerreviewmaketitle

\section{Introduction}

 In the emerging 5G radio systems, one of the envisioned key scenarios is a high-density network of low-bitrate terminals and sensors with applications, e.g., in remote metering, health care, smart homes and vehicles \cite{boccardi_five_2014, banelli_modulation_2014, gupta_survey_2015}. This use case, often referred to as the Internet-of-Things (IoT), targets to serve enormous numbers of devices, potentially up to 1 million devices per km$^2$, and hence 5G and beyond base-stations (BSs) should have extremely good spatial multiplexing and beamforming capabilities. Intense research efforts have recently suggested that one potential solution is to adopt very large antenna arrays, up to hundreds of antenna units, at the BSs together with proper transmitter and receiver processing, commonly known as massive MIMO \cite{rusek_scaling_2013, ngo_energy_2013}.

In networks with very large numbers of devices and BSs being equipped with large number of antenna units together with associated RF chains, the cost- and energy-efficiency (bit/Joule) aspects are particularly emphasized. Typically, the stringent requirements to reduce costs and energy consumption translate to the use of low-cost components with non-ideal characteristics. In this respect, there is an urgent need to address the problems regarding the achievable communication performance as well as required waveform schemes, transceiver architectures, and signal processing methods under such non-ideal hardware modules. 

In the context of multi-user communications, \cite{gustavsson_impact_2014, bjornson_massive_2015} provide a detailed analysis and discussion on the uplink rates achievable by massive MIMO BSs with non-ideal hardware. In \cite{pitarokoilis_uplink_2015}, the authors provide a sum-rate analysis for massive MIMO uplink impaired by phase-noise showing the imposed limitations on data transmission interval and number of scheduled users. {\color{black}Then focusing on a single-user scenario, \cite{mo_capacity_2015, nossek_2006, mezghani_ultra-wideband_2007} study the achievable MIMO downlink rates and mutual information under different assumptions regarding the channel state information (CSI) at transmitter (TX) and/or receiver (RX) and for high and low SNR regimes, respectively, when the receiver deploys 1-bit ADCs. However, no actual solutions or techniques for the waveform design or spatio-temporal precoding is proposed. Then, the article \cite{ivrlac_2006} considers space-time coding for improved link reliability in 1-bit quantized single-user MIMO systems while, again, no methods or solutions are proposed how beyond 1 bit/s/Hz spectral efficiency, per real domain and per RX antenna, can be achieved.} Regarding the signal processing and waveform aspects of fully digital beamforming based massive MIMO BSs, \cite{prabhu_hardware_2014, kammoun_linear_2014, zarei_low-complexity_2013} propose hardware-efficient implementations of linear spatial precoding which target for better energy-efficiency at massive MIMO transmitters. The works in \cite{mohammed_per-antenna_2013, mohammed_constant-envelope_2013} also focus on energy-efficiency of massive MIMO TXs and propose constant-envelope waveform design, enabling the use of energy-efficient non-linear power amplifiers in base-stations. 

{\color{black}Interestingly, based on the above literature review, the problem of \textit{multi-user waveform design and optimization to facilitate very low-cost and highly energy-efficient sensor receivers} is not yet addressed in the existing literature, especially in the timely IoT or Internet-of-Everything (IoE) use case where receivers can potentially adopt ADCs with very low, possibly even single-bit, conversion accuracy. In such IoT/IoE use case, where the number of user devices or terminals can be in the order of 1 billion or even more \cite{5g}, minimizing the energy-consumption, silicon area, and implementation costs of the terminals, through the adoption of highly simple hardware, is a primary concern.} To this end, in a very different context of high rate (multi-Gbps) single-input single-output (SISO) applications, \cite{landau_intersymbol-interference_2013, krone_communications_2012} provide a related study, seeking to enhance the energy-efficiency of the receiver by adopting simple 1-bit ADCs together with oversampling. In \cite{landau_intersymbol-interference_2013}, the mutual information analysis shows that it is in principle possible to achieve rates higher than 1 bit/s/Hz per real dimension even when adopting such 1-bit quantization (per real dimension) in the receiver. In the same work, a linear temporal precoding scheme in the form of controlled inter-symbol-interference (ISI) is proposed where a brute force method is used to optimize the overall ISI in the considered single-user SISO case. Then,\cite{gokceoglu_waveform_2016} extended the system model in \cite{landau_intersymbol-interference_2013} to frequency-flat block-fading MISO channels, and proposed well-structured and hence faster solutions to the ISI optimization problem compared to the approach in \cite{landau_intersymbol-interference_2013}.

\begin{figure*}[h]
	\centering
	\includegraphics[width=0.85\linewidth]{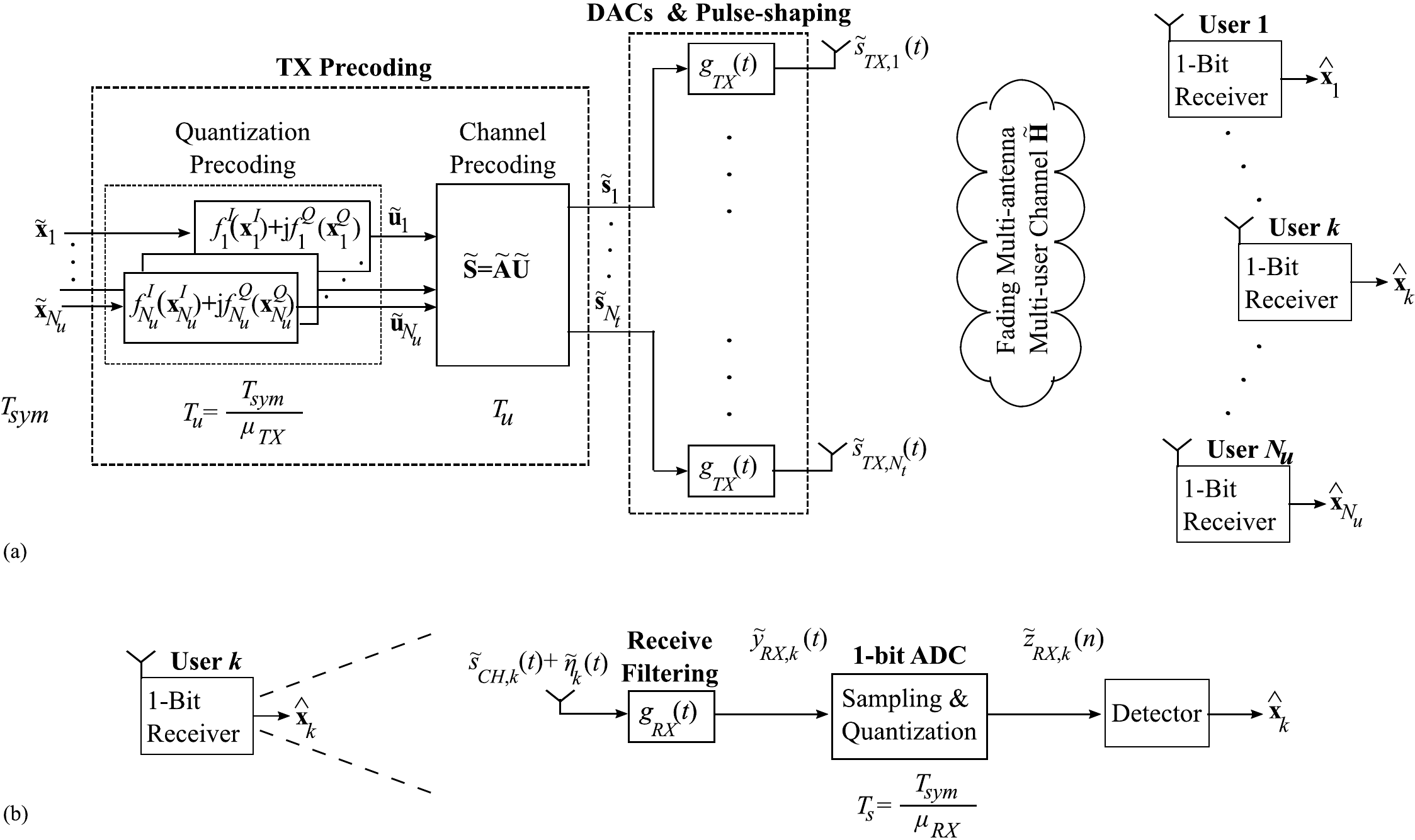} 
	\caption{(a) Considered massive MIMO downlink communication system with novel spatio-temporal precoding and waveform design, serving simultaneously $N_u$ users with 1-bit receivers. (b) More detailed illustration of the elementary receiver processing blocks. In general, all illustrations reflect complex-valued baseband equivalent notations, and also deliberately neglect different amplification stages in TX and RX chains for notational simplicity.}
	\label{fig:MIMOsystem1bit_TRX}
\end{figure*}

In this paper, we study a multi-user massive MIMO downlink system where the BS serves simultaneously multiple users, each equipped with simple 1-bit receivers (RXs), at the same time-frequency resource, as shown in  Fig. \ref{fig:MIMOsystem1bit_TRX}. \textcolor{black}{Our primary application scenario is related to very low-cost and highly energy-efficient IoT/IoE/sensor receivers, and hence we assume that the receivers are all single-antenna devices. This is also well in line with the recent commercial mobile cellular radio system standardization, in particular 3GPP narrowband IoT (NB-IoT), where the corresponding user equipment (UE) are single-antenna devices.\footnote{For more technical details related to 3GPP Narrowband IoT technology, please refer to the technical report TR 36.802 V13.0.0, available at http://www.3gpp.org/, and the Release 13 specifications TS 36.101 V13.4.0 and TS 36.211 V13.2.0.} There are, however, no technical limitations in the proposed solutions, specific to single-antenna receivers, and thus the developed techniques can be applied also in the context of multi-antenna receivers.} Specifically, we focus on the composite TX waveform design and optimization, in the form of novel spatio-temporal precoding, allowing the spatially multiplexed users to have spectral efficiencies higher than 1 bit/s/Hz (per real dimension, and per receiver antenna) despite the corresponding receivers preserve only 1-bit information of the received signal. The main rationale is to include the information about the transmitted symbol into the controlled ISI of the oversampled signal such that the receiver can extract this information and can uniquely decode the transmitted symbol using the 1-bit quantized observation. In this respect, we substantially extend the works in \cite{gokceoglu_waveform_2016} and \cite{landau_intersymbol-interference_2013} with several novel contributions which can be listed as follows: \vspace*{1mm} 

\hspace*{1mm} 1) In contrast to fairly simple single-user studies in \cite{gokceoglu_waveform_2016} and \cite{landau_intersymbol-interference_2013}, this paper focuses on multi-user massive MIMO scenario where multiple users, each equipped with a 1-bit receiver, are served simultaneously at the same time-frequency resource. 

\vspace*{0.5mm}
\hspace*{1mm} 2) In \cite{gokceoglu_waveform_2016} and \cite{landau_intersymbol-interference_2013}, the search or optimization of the ISI sequence is done for a simplified system model where there are no constraints on the TX and/or RX filters. However, practical TX filters have transmission bandwidth constraints stemming from the spectral mask and emission requirements of the system whereas practical RX filters have bandwidth constraints determined by the desired suppression level of adjacent channel signals and other out-of-band noise and interference. In this paper, we consider such practical TX and RX filters which both contribute to the overall observable ISI in the receivers. 

{\color{black}
\vspace*{0.5mm}
\hspace*{1mm} 3) The overall quantization precoder design is formulated as a constrained convex optimization problem, seeking to maximize the robustness against noise and residual IUI, under the practical constraints related to bandlimitation, TX power and TX/RX pulse-shape filtering.
}

\vspace*{0.5mm}
\hspace*{1mm} 4) Furthermore, the problem formulation is also largely generalized to facilitate arbitrary ISI memory lengths unlike the studies of \cite{gokceoglu_waveform_2016} and \cite{landau_intersymbol-interference_2013} where the ISI memory is strictly limited to 1 symbol duration. 

\vspace*{0.5mm}
\hspace*{1mm} 5) Finally, in contrast to \cite{gokceoglu_waveform_2016} and \cite{landau_intersymbol-interference_2013}, the precoding function of input-symbols is not limited to be linear in this paper. We show that the direct design of the precoding function is challenging for arbitrary lengths of ISI memory and hence, alternatively, we propose designing and optimizing the processed (precoded) output samples directly.  \vspace*{0.5mm} 

The rest of this paper is organized as follows. The basic system and signal models are introduced in Section \ref{sec:Basic System and Signal Models}. Then, the principles of multi-user-multiantenna waveform design and detection are addressed in Section \ref{sec:Principles of Spatio-temporal Waveform Design and Detection for 1-bit Receivers}. The actual design of quantization precoded sequence in the form of a constrained convex optimization problem is carried out in Section \ref{sec:Design of Quantization Precoded Sequence}. In Section \ref{sec:Simulations}, numerical evaluations are provided to illustrate the symbol-error-rate (SER) performance of the system with respect to various system parameters. Finally, conclusions are drawn in Section \ref{sec:Conclusion}. Derivation of the received signal model is given in Appendix A, while the constraints used in the optimization problem formulation are derived in Appendix B. 

Throughout the paper, complex-valued scalars, vectors and matrices are indicated with a tilde sign on top, i.e. $\tilde{x}$, $\tilde{\bf x}$ and $\tilde{\bf X}$. The matrix ${\bf I}_{a}$ denotes an $a \times a$ identity matrix whereas ${\bf 0}_{a \times b}$, ${\bf 1}_{a \times b}$ denote $a \times b$ matrices of all 0's and 1's, respectively. Absolute value of a real-valued vector $\mathbf{x}$ is computed element-wise, and is not to be confused with vector norm. The $\text{diag}\left(.\right)$ operator returns a diagonal matrix with diagonal entries being the elements of the argument vector.
The function $\text{sign}(.)$ returns the sign of the real-valued scalar in its argument, i.e., $\text{sign}(a)=1$ if $a\ge 0$ and $\text{sign}(a)=-1$ if $a<0$ whereas the sign is computed element-wise for vector-valued arguments. The linear convolution is denoted with $\otimes$.

\section{Basic System and Signal Models} \label{sec:Basic System and Signal Models}

\subsection{Preliminaries}

In this article, we consider a time-division duplex (TDD) based multi-user massive MIMO network, and focus on its downlink where BS has $N_t$ antennas and serves $N_u$ receivers simultaneously at a given time-frequency resource, as illustrated in Fig. 1. {\color{black}Stemming from the assumed IoT/IoE/sensor application and to simplify the notations, the receivers are all assumed to be single-antenna devices. Such single-antenna user device assumption is also fairly common in the existing massive MIMO literature, such as the seminal works \cite{rusek_scaling_2013, ngo_energy_2013, bjornson_massive_2015}. There are, however, no technical limitations in applying the developed techniques also in the context of multi-antenna UE receivers in massive MIMO networks, as long as the spatial channels from the base-station array towards the individual antennas of a given UE are mutually uncorrelated.} The receivers are assumed to have signal observations with only 1-bit/real-dimension quantization and an integer oversampling factor of $\mu_{RX}>1$ with respect to symbol rate $T_{sym}$, i.e., receiver sampling period is $T_s=\frac{T_{sym}}{\mu_{RX}}$. For each user, the target is to achieve a spectral efficiency of $1<R<\text{log}_2(R_{in})$ bit/s/Hz per real-dimension where $R_{in}$ is the input modulation size per real-dimension. For a given transmission bandwidth of $W_{TX}$ and input modulation size $R_{in}$, the spectral efficiency per real-dimension is given by $R=\frac{\text{log}_2(R_{in})}{T_{sym}W_{TX}}$ bits/s/Hz. By transmission bandwidth $W_{TX}$, we mean that the baseband equivalent transmit waveform is essentially bandlimited to frequencies of $-\frac{W_{TX}}{2} \cdots \frac{W_{TX}}{2} $, in two-sided spectral notations. Then following directly from the above spectral efficiency constraints, the transmission bandwidth should satisfy 
\begin{equation} \label{eq:wtx_range}
\frac{1}{T_{sym}}<W_{TX}<\frac{\text{log}_2(R_{in})}{T_{sym}}
\end{equation}
where the upper limit corresponds to $R=1$.

The basic baseband equivalent block-diagram of the assumed communication system with essential TX and RX processing blocks is shown in Fig. \ref{fig:MIMOsystem1bit_TRX}. In the basic system model, the $i^{th}$ complex data symbol to be transmitted to user $k$ is denoted as $\tilde{x}_k(i)=x^I_k(i)+jx^Q_k(i)$. It is assumed that both the real and imaginary parts take values from the set $ X_{in}=\{b_1,b_2, \cdots, b_{R_{in}}\}$ with size or cardinality $\mid X_{in}\mid=R_{in}$, i.e., for $ \theta \in \{I,Q\}$, $x^{\theta}_k(n)\in X_{in}$. The corresponding symbol vector is denoted by $\tilde{\bf x}_k={\bf{x}}^I_k+j{\bf{x}}^Q_k$, where ${\bf{x}}^{\theta}_k=[x^{\theta}_k(0), x^{\theta}_k(1), \cdots, x^{\theta}_{k}(N_{block}-1)]^T$ and $N_{block}$ denotes the number of symbols to be communicated to user $k$. Then the composite block of $N_u \times N_{block}$ symbols to be communicated to all simultaneously scheduled users is denoted with the matrix $\tilde{\bf X}={\bf{X}}^I+j{\bf{X}}^Q$ where ${\bf{X}}^{\theta}=\begin{bmatrix}{\bf{x}}^{\theta}_1, \cdots, {\bf{x}}^{\theta}_{N_u} \end{bmatrix}^T$.

\subsection{Spatio-Temporal Precoding}

The composite input-symbol block is processed by TX precoder which, at a general level, maps the matrix of original data-symbols $\tilde{\bf X}$ to a matrix of precoded samples $\tilde{\bf S}$. In this work, we consider a special case of this mapping which is done in two stages, namely the novel quantization precoding followed by a maximum-ratio transmission (MRT) or zero-forcing (ZF) type spatial channel precoder. The role of the channel precoder is to provide coherent combining of a particular user stream at the intended RX whereas the other user streams are combined incoherently and thereby provide separation of user streams together with beamforming gain in desired signal strength. On the other hand, \textit{the role of the novel quantization precoding is to enable a particular RX to decode its intended stream considering that the RX has oversampled 1-bit observations which carry only the sign information of the underlying signal.} This decomposition separates the two problems, namely 1) spatial multiplexing of multiple user streams, and 2) temporal coding of each user stream enabling the intended 1-bit RX to uniquely decode the input-symbols from quantized observations. Such two-stage decomposed spatio-temporal precoding approach is directly stemming from the physical propagation and processing chains, namely the transmitted antenna signals are first propagating through the multi-user MIMO channel, after which the observation of a given receiver is being sampled and quantized by the single-bit ADC. Thus, the overall waveform optimization building on the spatio-temporal two-stage precoding approach is designed in a corresponding manner, such that channel precoding matches the propagation channel while the quantization precoding matches then the single-bit ADC characteristics. 

In this two-stage processing, the quantization precoding first maps the data-matrix $\tilde{\bf X}$ to a precoded matrix $\tilde{\bf U}$ expressed as
\begin{equation} \label{eq:Umat}
\tilde{\bf U}=\begin{bmatrix}\tilde{\bf{u}}^T_1 \\ \vdots \\ \tilde{\bf{u}}^T_{N_u} \end{bmatrix}=\begin{bmatrix}\left(f^I_1({\bf{x}}^I_1)+jf^Q_1({\bf{x}}^Q_1)\right)^T \\ \vdots \\ \left(f^I_{N_u}({\bf{x}}^I_{N_u})+jf^Q_{N_u}({\bf{x}}^Q_{N_u})\right)^T \end{bmatrix}
\end{equation} 
where $\tilde{\bf u}_k={\bf u}_k^I+j{\bf u}_k^Q$ with ${\bf u}_k^{\theta}=[{u}_k^{\theta}(0), \cdots, {u}_k^{\theta}(N_q-1)]^T$. Quantization precoded symbols are obtained by processing the $I$ and $Q$ branch data of each user stream separately, i.e., ${\bf u}_k^{\theta}= f^{\theta}_k({\bf x}_k^{\theta})$ where $ f^{\theta}_k: X_{in}^{N_{block}}\rightarrow \mathbb{R}^{N_q }$, and $N_q=\mu_{TX}N_{block}$ is the length of the quantization precoded sequence while $\mu_{TX}$ denotes the precoding oversampling factor. Separate per $I$ and $Q$ quantization precoding is directly stemming from the fact that the RX side single-bit ADCs also operate separately on the $I$ and $Q$ components of the received signal. In this work we mainly study two choices of $\mu_{TX}$, namely $\mu_{TX}=1$ and $\mu_{TX}=\mu_{RX}$. The former one corresponds to mapping from input-symbols to a sequence of samples at input symbol-rate whereas in the latter case, the sequence of precoded samples appears at the same oversampled rate as in the RX. In general, the sample period after the quantization precoding stage can be written as $T_u=\frac{T_{sym}}{\mu_{TX}}$. Detailed descriptions of novel quantization precoding design and optimization are provided in Sections \ref{sec:Principles of Spatio-temporal Waveform Design and Detection for 1-bit Receivers} and \ref{sec:Design of Quantization Precoded Sequence}, while here the presentation is limited to general-level input-output expressions to establish fundamental signal models that are then utilized in the actual design and optimization. 

Quantization precoding is followed by precoding via an $N_t \times N_u$ spatial channel precoder matrix $\tilde{\bf{A}}$, written here as
\begin{equation} \label{eq:S_mat}
\tilde{\bf S}=\tilde{\bf{A}}\tilde{\bf U}={\bf{S}}^I+j{\bf{S}}^Q
 \end{equation}
where ${\bf{S}}^{I}=\text{Re}\{\tilde{\bf{A}}\tilde{\bf U}\}=\begin{bmatrix}{\bf{s}}^{I}_1, \cdots, {\bf{s}}^{I}_{N_t} \end{bmatrix}^T$, ${\bf{S}}^{Q}=\text{Im}\{\tilde{\bf{A}}\tilde{\bf U}\}=\begin{bmatrix}{\bf{s}}^{Q}_1, \cdots, {\bf{s}}^{Q}_{N_t} \end{bmatrix}^T$ and ${\bf{s}}_k^{\theta}=[s_k^{\theta}(0), \cdots, s_k^{\theta}(N_q-1)]^T$. Throughout this paper, we assume narrowband flat fading channels between the BS transmitter and $N_u$ receivers, collected into an $N_u \times N_t$ multi-user MIMO channel matrix $\tilde{\bf{H}}$ where $\tilde{\bf{H}}_{k,l}=\tilde{h}_{k,l}$ is the complex fading coefficient from $l^{th}$ TX antenna to $k^{th}$ user RX having i.i.d. zero-mean Gaussian distribution with variance $\sigma^2_h$ where without of loss of generality we set $\sigma^2_h=1$. {\color{black}The narrowband assumption is well-justified in the considered IoT/IoE/sensor applications where commonly the radio access channel bandwidth is only in the order of a few tens to a few hundreds of kHz, and thus with any reasonable delay spread, especially at the center-frequencies where massive MIMO is feasible, the channel is essentially flat fading.} Then, the corresponding ZF and MRT spatial precoders are given up to a scaling constant as \cite{ngo_massive_2013, yang_performance_2013}
\begin{equation} \label{eq:MRTZF}
 \tilde{\bf A}=\begin{cases} c_{MRT}\hat{\bf H}^H, &\text{for MRT} \\ 
	c_{ZF}\hat{\bf H}^H(\hat{\bf H}\hat{\bf H}^H)^{-1}, &\text{for ZF} \end{cases}
\end{equation}
where $\hat{\bf H}$ refers to the corresponding estimated multi-user MIMO channel, while \textcolor{black}{the scaling aspects and their relation to chosen transmit power constraints are discussed in detail in Section IV-A and Appendix \ref{app:constraints}}. 

In TDD systems, stemming from channel reciprocity assumption, such downlink channel state information (CSI) can be obtained using, e.g., orthogonal uplink pilots \cite{marzetta_how_2006, ngo_massive_2013}. \textcolor{black}{We note that the spatial channel precoder processes the quantization precoder output samples whose rate is, in general, $\mu_{TX}$ times the symbol rate $1/T_{sym}$. Thus, compared to ordinary symbol-rate spatial precoding, the processing complexity is increased but only moderately so for reasonable values of  $\mu_{TX}$, such as $\mu_{TX}=2$.} {\color{black}In general, the MRT and ZF precoders are well-established and widely-adopted spatial processing solutions in the existing massive MIMO literature. There are, however, no prior works in the open scientific literature where such spatial precoders would be complemented with novel temporal precoding, in the overall space-time array processing, such that 1-bit/real-dimension receivers can be adopted.}

\subsection{TX/RX filtering and 1-bit Quantized Received Signal Model}

The discrete-time samples at the output of the overall TX precoder are converted to continuous waveforms at the $I$ and $Q$ branches of each antenna unit via digital-to-analog converters (DACs) and TX filters. The transmitted continuous waveforms from each antenna branch pass then through multi-user MIMO channel arriving at the receivers. The $I$ and $Q$ components of the noisy received signal are then filtered via identical real-valued low-pass receive filters which are essentially limiting the received signal double-sided bandwidth to the useful signal bandwidth $W_{TX}$ and suppressing all spectral components outside this band. Then, the filtered $I$ and $Q$ signals are converted to digital samples via 1-bit ADCs which are effectively modeled as sampling and single-bit quantization. Notice that typically BSs and also high-throughput terminals with good processing power adopt both analog and (multi-rate) digital filters for channel selection whereas here the channel selectivity is assumed to be achieved purely through analog RX filters. This complies well with the target scenario of highly simple IoT/sensor receivers, addressed in this article.

Appendix \ref{app:TRX_process} provides detailed derivations and discussions related to more detailed TX and RX filtering characteristics, channel hardening principle and the corresponding signals at different stages. Based on those derivations, the sampled complex noisy received signal at user $k$, prior to 1-bit quantization, can be expressed as
\begin{align}\label{eq:sRX_m_text}
\tilde{y}_{RX,k}(n) &=\tilde{s}_{RX,k}(n)+\tilde{\eta}_{filt,k}(n) \nonumber \\
			&={y}^I_{RX,k}(n)+j{y}^Q_{RX,k}(n)
\end{align}
where ${y}^{\theta}_{RX,k}(n)={s}^{\theta}_{RX,k}(n)+{\eta}^{\theta}_{filt,k}(n)$ with ${\eta}^{\theta}_{filt,k}(n)$ denoting the filtered Gaussian noise process. The useful signal term, following from the last line of \eqref{eq:sRX_sample}, can be expressed as
\begin{equation} \label{eq:sRX_sample_theta}
{s}^{\theta}_{RX,k}(n)=\beta_k {\bf g}^T_{tot,k}(n){\bf u}^{\theta}_k
\end{equation}
where $\beta_k>0$ is the real-valued beamforming gain and ${\bf g}_{tot,k}(n)$ is the effective filtering vector stemming from TX and RX pulse-shape filtering which is explicitly given in Appendix \ref{app:TRX_process}. Notice that \eqref{eq:sRX_sample_theta} gives the received signal model as a convolution between the quantization precoded symbols and this effective filter resulting from the TX and RX pulse-shaping, and for generality allows also the possibility of different oversampling factors and pulse bandwidths in the TX and RX. 

The final outputs of the ADCs in the $I$ and $Q$ branches are then formed by the single-bit quantization of the sampled received signal, and can be compactly written as
\begin{equation}\label{eq:quantized_i}
	{z}^{\theta}_{RX,k}(n)=\text{sign}(y^{\theta}_{RX,k}(n)).
\end{equation}
{\color{black}Notice that while the fundamental signal models (5)-(7) are here derived for single-antenna UEs, exactly the same models can be used in the context of multi-antenna UE devices, assuming that the spatial channels from the base-station array towards the individual antennas of a given UE are sufficiently uncorrelated. In an extreme example, all $N_u$ receive antennas can belong to the same UE and thus the user index $k$ corresponds essentially to the receiver antenna index towards which the particular transmit stream is beamformed. To this end, spatial multiplexing of multiple streams towards a single UE and the waveform optimization solutions developed in this article are complementary techniques to increase the spectral efficiency of the radio link and the overall radio system under 1-bit receivers.}

\subsection{Detection}

In the final stage of RX processing, based on the 1-bit quantized $I$ and $Q$ observations, the receiver performs detection, i.e., makes decisions about the transmitted input symbols. Here, stemming from the fact that the terminals and/or sensors are assumed to be very simple devices, i.e., have only limited signal processing capabilities, we focus on quantization precoding schemes that allow very \textit{low-complexity symbol-detection}. Given that, the detected $i$'th input-symbol $\hat{x}_k(i)=\hat{x}^I_k(i)+\hat{x}^Q_k(i)$ at user $k$ receiver is obtained through a memoryless detector defined formally as
\begin{equation} \label{eq:detector_main}
	\hat{x}^{\theta}_k(i)=d^{\theta}_k({\bf z}^{\theta}_{RX,k}(i)) 
\end{equation}
where $d^{\theta}_k:\mathbb{R}^{\mu_{RX}}\rightarrow X_{in}$ is the detection function and the $\mu_{RX} \times 1$ vector of quantized observations used by the detector to make a decision on ${x}^{\theta}_k(i)$ is given as
\begin{equation}\label{eq:quantized_z_i}
	{\bf z}^{\theta}_{RX,k}(i)=[z^{\theta}_{RX,k}(N_i), \cdots, z^{\theta}_{RX,k}(N_i+\mu_{RX}-1)]^T
\end{equation}
with $N_i$ being the starting sample index that corresponds to input symbol $i$. The term memoryless refers here to the fact that only the 1-bit samples of the given symbol duration are used in the detection of the corresponding data symbol. Notice also that we have specifically formulated the detection function separately for the $I$ and $Q$ branches, which is applicable for square-QAM symbol constellations.

The quantized received vector is written as ${\bf z}^{\theta}_{RX,k}(i)=\text{sign}({\bf{y}}^{\theta}_{RX,k}(i))=\text{sign}({\bf{s}}^{\theta}_{RX,k}(i)+\boldsymbol{\eta}^{\theta}_{filt,k}(i))$. The noise vector is given as $\boldsymbol{\eta}^{\theta}_{filt,k}(i)=[{\eta}^{\theta}_{filt,k}(N_i), \cdots, {\eta}^{\theta}_{filt,k}(N_i+\mu_{RX}-1)]^T$ whereas the unquantized signal vector ${\bf{s}}^{\theta}_{RX,k}(i)=[{s}^{\theta}_{RX,k}(N_i), \cdots, {s}^{\theta}_{RX,k}(N_i+\mu_{RX}-1)]^T$ can be expressed based on \eqref{eq:sRX_sample_theta} as
\begin{equation} \label{eq:sRX_vec_det}
	{\bf{s}}^{\theta}_{RX,k}(i)=\beta_k{\bf G}_{tot,k}(i){\bf u}^{\theta}_k
\end{equation}
where ${\bf G}_{tot,k}(i)=[{\bf g}_{tot,k}(N_i), \cdots, {\bf g}_{tot,k}(N_i+\mu_{RX}-1)]^T$. 
{\color{black}In the case of a multi-antenna receiver, the detection of the multiple streams is carried out in a per-receiver-antenna manner, that is, for the applicable range of $k$'s in parallel.}

Detailed developments related to the form of the detector function $d^{\theta}_k$ and other implementation details are provided in Section \ref{sec:unique_detection}. Notice that the ISI in \eqref{eq:sRX_vec_det} is fundamentally of continuous sliding-window nature, creating memory and dependence between neighboring samples. Classically, this is modeled through state machine and the detection is addressed using Viterbi algorithm based sequence detector. In our case, however, the ISI appearing in the received signal model \eqref{eq:sRX_vec_det} is between the \textit{precoder output samples}, which are directly influenced and controlled in the waveform optimization. Thus, as long as the precoder optimization is carried out properly, reliable detection can be achieved using a simple memoryless symbol level detector, using only the 1-bit quantized received samples of the corresponding symbol duration. This will be elaborated in more details in the upcoming sections.


\section{Principles of Waveform Design and Detection} \label{sec:Principles of Spatio-temporal Waveform Design and Detection for 1-bit Receivers}

In this section, stemming from the previous signal and system models, we discuss the main objectives and optimization criteria for the spatio-temporal precoding based waveform design at TX together with the corresponding detection processing using the oversampled 1-bit quantized observations at RX.  

\subsection{Objectives of the Waveform Design}

Given an RX oversampling factor $\mu_{RX}$, the input data block $\tilde{\bf X}$, as well as the TX and RX filters, i.e., $g_{TX}(t)$ and $g_{RX,k}(t)$  $\forall k$, the objectives of the quantization precoding and detection functions, $f^{\theta}_k$ and $d^{\theta}_k$ $\forall k$, are defined as follows: 
\vspace*{1.5mm}

Objective 1) In the absence of noise and IUI, user $k$ can uniquely detect the intended input-symbol ${{x}}^{\theta}_k(i)$, i.e., ${{\hat{x}}}^{\theta}_k(i)={{x}}^{\theta}_k(i)$, using only ${\bf z}^{\theta}_{RX,k}(i)$, i.e., only the received quantized 1-bit samples of the corresponding data symbol period, as implied in \eqref{eq:detector_main}-\eqref{eq:quantized_z_i}. 
\vspace*{1.5mm}

Objective 2) In the presence of noise and residual IUI, all users have good symbol-error rate (SER) performance. In other words, there is built-in robustness against noise and residual IUI.
\vspace*{0.1mm} \\

We address the quantization precoder design such that the precoder outputs ${\bf{u}}^I_k$ and ${\bf{u}}^Q_k$ are directly optimized for given input streams ${\bf{x}}^I_k$ and ${\bf{x}}^Q_k$. Notice that the model for the received signal prior to quantization, i.e., ${s}^{\theta}_{RX,k}(n)=\beta_k {\bf g}^T_{tot,k}(n){\bf u}^{\theta}_k$ in \eqref{eq:sRX_sample_theta}, indicates that both ${\bf{u}}^I_k$ and ${\bf{u}}^Q_k$ are processed identically with ${\bf g}^T_{tot,k}(n)$. Moreover, the filtered noise ${\eta}^{\theta}_{filt,k}(n)$ has identical statistical characteristics for all $\theta \in {I,Q}$. Due to such identical processing and statistical characteristics, without loss of generality, we will address only the design of ${\bf{u}}^I_k$ while the design of ${\bf{u}}^Q_k$ follows similarly. For notational simplicity, from this point on, we also drop the superscript $I$ and hence we have the variable and function substitutions ${\bf{x}}_k \leftarrow {\bf{x}}^I_k$, ${\bf{u}}_k \leftarrow {\bf{u}}^I_k$, $d_k \leftarrow d^{I}_k$ and so forth.

\subsection{Criterion for Unique Detection in Noiseless Scenario and Detection under Noise} \label{sec:unique_detection}

Under the assumptions of Objective 1) above, the following properties are sufficient to enable unique detection of input-symbols in the absence of noise.
\vspace*{1.5mm} 

\hspace*{2mm}  P1) First, the RX oversampling factor needs to be selected such that $\mu_{RX} \ge \log_{2}(R_{in})$. There are then altogether $R_{out}=2^{\mu_{RX}}$ different possible quantized outputs or codewords and the corresponding codeword set is denoted by $C_{out}=\{{\bf c}_1, {\bf c}_2, \cdots, {\bf c}_{R_{out}}\}$ where every element of any codeword belongs to set $\{-1, +1\}$. 
\vspace*{1.5mm}

\begin{figure}
    	\centering
        	\includegraphics[width=\linewidth]{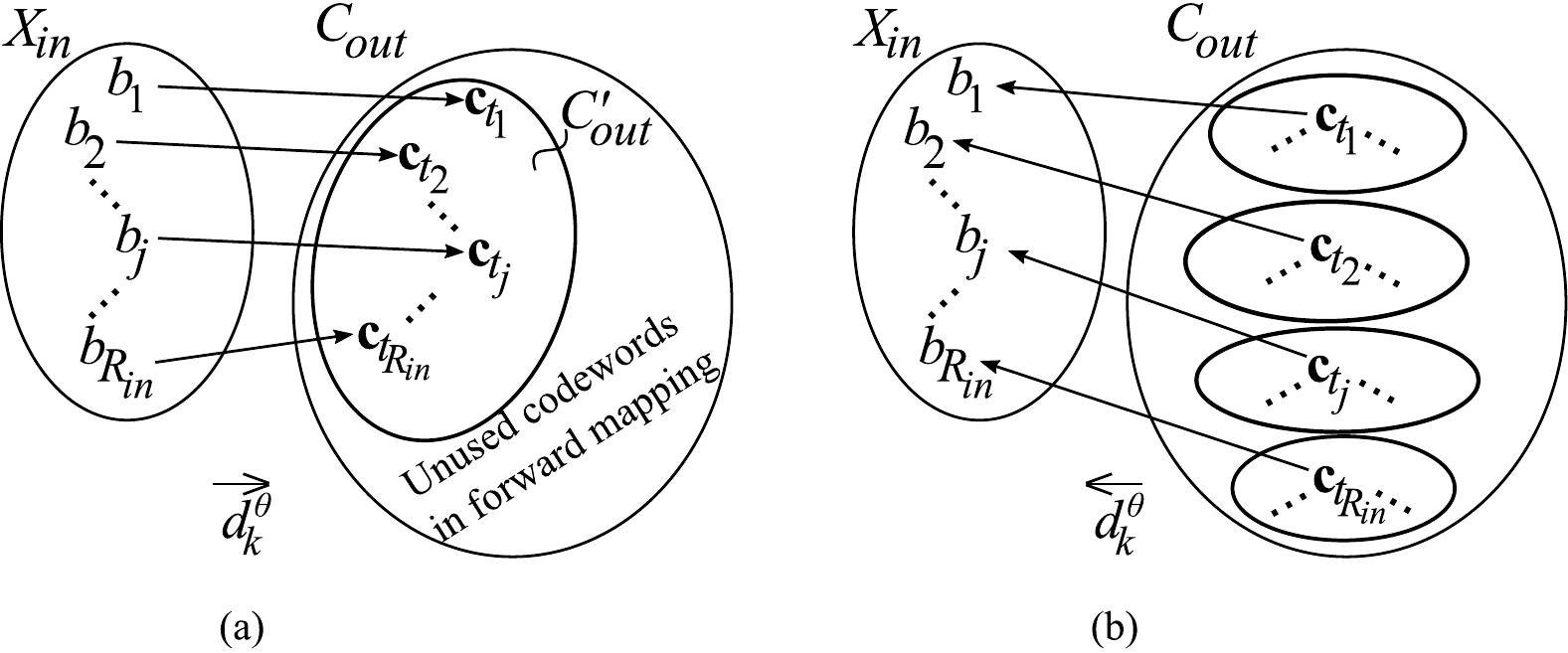} 
	\caption{Illustration (a) depicts the forward mapping $\tightoverset{\rightarrow}{d_k}$ of elements of input symbol set $X_{in}$ to elements of $C'_{out}$ which is a subset of the overall codeword set $C_{out}$. In (b), the corresponding backward mapping $\tightoverset{\leftarrow}{d_k}$ from $C'_{out}$ to $X_{in}$ is shown, together with spheres of words that are closest in Hamming distance sense to the valid codewords. }
	\label{fig:mapping}
\end{figure}

\hspace*{2mm} P2) A one-to-one forward mapping \textcolor{black}{(FM)} is defined from $X_{in}$ to subset of $C_{out}$, referred to as $C'_{out}$, where $C'_{out} \subseteq C_{out}$ such that $\tightoverset{\rightarrow}{d_k}:X_{in}\rightarrow C'_{out}$, as shown in Fig. \ref{fig:mapping} (a). Hence, for each $1\le j \le R_{in}$ the  input symbol $b_j$ is mapped to a distinct codeword ${\bf c}_{t_j} \in C'_{out}$ with $1\le t_j \le R_{out}$, i.e., $b_1 \rightarrow {\bf c}_{t_1}, \cdots, b_{R_{in}} \rightarrow {\bf c}_{t_{R_{in}}}$. The number of possible \textcolor{black}{FM}s follows from straightforward combinatorics as $N_{map}=R_{out}\times (R_{out}-1) \times \cdots (R_{out}- R_{in}+1)= {R_{out} \choose R_{in}}R_{in}!$.
\vspace*{1.5mm}

\hspace*{2mm} P3) A sequence ${\bf{u}}_k$ is searched such that at each symbol interval $i$, in the absence of noise and interference, the quantized received vector is a unique codeword of $C'_{out}$, corresponding to the input symbol, i.e., $\forall i$ if $x_k(i)=b_{j}$, then ${\bf{z}}_{RX,k}(i)={\bf c}_{t_j}$. 
\vspace*{1.5mm}

\hspace*{2mm} P4) The detection is performed building on a backward mapping $\tightoverset{\leftarrow}{d_k}:C'_{out} \rightarrow X_{in} $ which is illustrated in Fig. \ref{fig:mapping} (b). 
In the absence of noise and given P1)-P3), the detection can simply be performed by the inverse of the \textcolor{black}{FM}, i.e., $ {\bf c}_{t_1}  \rightarrow b_1, \cdots, {\bf c}_{t_{R_{in}}} \rightarrow  b_{R_{in}} $. \vspace*{0.1mm} \\

The above properties establish sufficient conditions for unique detection in the noiseless case. Next, we generalize the detection processing to the practical case of noisy reception, still assuming $R_{out}>R_{in}$. Stemming from the previous signal models and the codeword interpretations, the overall symbol detector, denoted as $d_k$, compares the received quantized signal vector during a symbol interval $i$, namely ${\bf{z}}_{RX,k}(i)$, to all valid codewords in $C'_{out}$ using Hamming-distance metric and the one with minimum distance is selected. Then, the input symbol is decoded via the backward mapping $\tightoverset{\leftarrow}{d_k}$. Formally, the $i$'th decoded symbol $\hat{x}_k(i)=d_k({\bf z}_{RX,k}(i))$ can now be defined as
\begin{equation}
\hat{x}_k(i)=\tightoverset{\leftarrow}{d_k}({\bf c}') , \quad \text{where} \quad {\bf c}': \min_{{\bf c}\in C'_{out}} \text{Hamming}({\bf{z}}_{RX,k}(i),{\bf c})
\end{equation}

{\color{black}Notice that the above detector is a low-complexity memoryless symbol-detector which is well in-line with the general target of low-cost and simple IoT/IoE/sensor receivers, pursued in this article.} Moreover, the detector has also low latency, since in order to detect the input symbol $x_k(i)$, only the quantized sample vector corresponding to this symbol interval, i.e., ${\bf{z}}_{RX,k}(i)$, is required.

\subsection{Criterion for Built-in Protection against Noise/IUI}
In general, the receiver thermal noise and residual IUI distort the useful received samples prior to quantization corresponding to input-symbol $i$, namely ${\bf{s}}_{RX,k}(i)$. For the case with $R_{out}=R_{in}$, the detection will be correct even under such distortion if the quantized vector ${\bf{z}}_{RX,k}(i)$ is not altered with respect to the true codeword ${\bf c}$, i.e., $\text{sign}({\bf{s}}_{RX,k}(i)+\boldsymbol{\eta}_{filt,k}(i))=\text{sign}({\bf{s}}_{RX,k}(i))={\bf c}$. Then in the case with $R_{out}>R_{in}$, the detection will be correct as long as the quantized output is still closer in Hamming distance sense to the true codeword ${\bf c}$ than to any other vector of $C'_{out}$.

In general, the noise and IUI are most likely to alter the sign of those unquantized signal samples that have a small amplitude. In this respect, an intuitive criterion to achieve Objective 2, through precoder optimization, is to maximize the minimum absolute value of the received signal samples prior to quantization. In other words, we try to maximize $\gamma_k$ such that \textcolor{black}{$\mid {\bf{s}}_{RX,k} \mid \succeq \gamma_k$ where $\succeq$} is an element-wise comparison of the left-hand side vector with right-hand side scalar and ${\bf{s}}_{RX,k} $ reads
\begin{equation} \label{eq:general_max}
	{\bf{s}}_{RX,k}= \beta_{k}{\bf G}_{tot,k} {\bf u}_k
\end{equation}
where ${\bf{s}}_{RX,k}$ is obtained by stacking ${\bf{s}}_{RX,k}(i)$ over the processed symbol block, i.e. $1\le i \le N_{block}$ exluding the starting and ending transients of the filters.

\begin{figure}
    	\centering
        	\includegraphics[width=\linewidth]{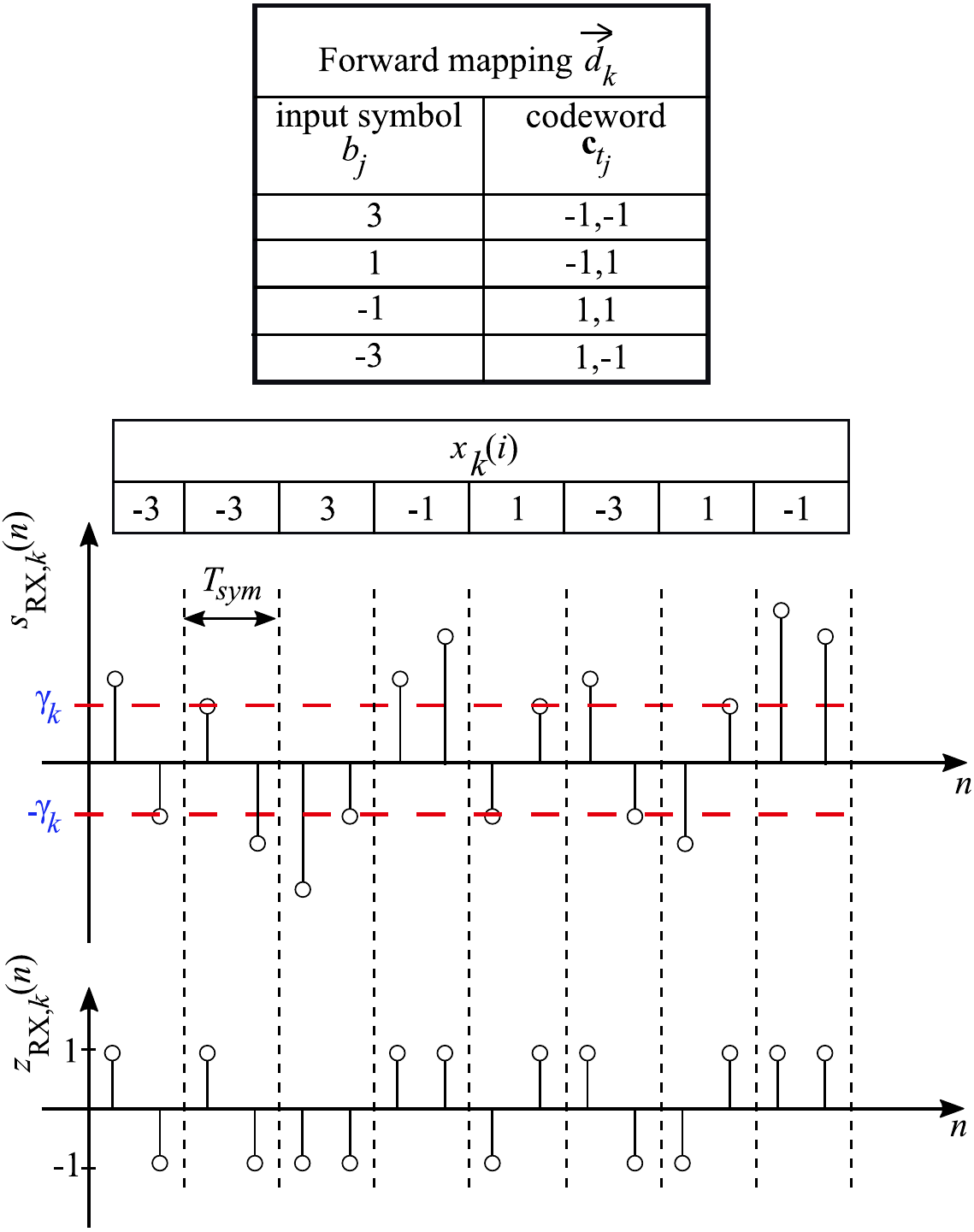} 
	\caption{Example input-symbol sequence $x_k(i)$ from 16-QAM input modulation (showing only the corresponding 4-ASK scheme per real-dimension)  and corresponding received signal samples with $\mu_{RX}=2$ before and after 1-bit quantization at each symbol of duration of $T_{sym}$. The \textcolor{black}{FM} $\tightoverset{\rightarrow}{d_k}$ is given in the top table and the minimum amplitude value $\gamma_k$ is also shown.}
	\label{fig:precoded_sequence}
\end{figure}

In Fig. \ref{fig:precoded_sequence}, an example input-symbol sequence from 16-QAM input modulation (showing only the corresponding 4-ASK scheme per real-dimension) and corresponding received signal samples at each symbol interval before and after quantization are illustrated. The FM $\tightoverset{\rightarrow}{d_k}$ between input-symbols and codewords is given in the table at the top of the figure and the minimum unquantized amplitude value $\gamma_k$ which is to be maximized is also shown. Note that for both input-symbols $x_k(0)$ and $x_k(1)$ which have the same value of $-3$, the corresponding unquantized signal samples, $s_{RX,k}(n)$, are generally not identical but have identical sign values. Hence, the corresponding quantized signal samples, $z_{RX,k}(n)$, are identical in the absence of noise and interference allowing the RX to uniquely detect the input-symbols based on the given mapping. Then, in the presence of noise and/or interference, $\gamma_k$ determines whether the sign of the (un)quantized samples will change or not. Consequently, this makes the difference between correct and erroneous detection. {\color{black}Notice that since maximizing $\gamma_k$ provides robustness also against IUI, which is generally stemming from the lack of spatial precoder being able to fully orthogonalize the parallel transmit streams, one can see that the optimization of the temporal quantization precoder can also implicitly contribute to the IUI suppression at the output of the 1-bit quantizers in the receivers.}

{\color{black}In general, it is important to note that the value of $\gamma_k$ depends on the desired signal gain $\beta_k$, provided by the spatial channel precoder, relative to noise and interference. Moreover, the channel precoder characteristics impact directly the amount of residual IUI. Thereby, we see a direct interaction between the two precoders where the achievable $\gamma_k$ and therefore the SER performance jointly depend on both quantization and channel precoders.}

\section{Optimization of Quantization Precoded Symbols} \label{sec:Design of Quantization Precoded Sequence}

In this section, building on the derived signal and system models, we address in detail the optimization of the quantization precoded vectors, namely ${\bf{u}}_k$. The design is constrained by the total BS transmission power denoted by $P_0$ and the final transmission bandwidth of each antenna signal denoted as $W_{TX}$. These constraints and their relations to TX pulse-shape filter characteristics, number of antennas, and the emission requirements are further elaborated in Appendix \ref{app:constraints}.

\subsection{Optimization Algorithm based on Maximizing the Minimum Absolute Value of Unquantized Received Samples }

In order to meet the Objective 2 stated in the previous section, the cost function to be maximized is here defined as the minimum absolute value of the total received signal vector given in \eqref{eq:general_max}. Now, the element-wise absolute value of this vector can be written as 
\begin{equation} \label{eq:abs_s}
	|{\bf{s}}_{RX,k}|={\bf C}_k {\bf{s}}_{RX,k}=\beta_k{\bf C}_k {\bf G}^{tot}_k {\bf u}_k
\end{equation}
where the codewords are chosen based on a \textcolor{black}{FM} satisfying unique detectability condition while the codeword matrix ${\bf C}_k$ is given by
\begin{equation} \label{eq:codeword_matrix} 
{\bf C}_k =\text{diag}\left([{\bf c}^T_{1}, \cdots, {\bf c}^T_{N_{block}}]^T\right).
\end{equation}
Notice that the first equality in \eqref{eq:abs_s} follows directly from the fact that the elements of the codewords are $+1$'s and $-1$'s, depending directly on the sign of the samples of vector ${\bf s}_{RX,k}$. Notice also that since $\beta_k>0$ is a positive scalar, maximizing the minimum element of expression \eqref{eq:abs_s} is equivalent to maximizing the minimum element of ${\bf C}_k {\bf G}_{tot,k} {\bf u}_k$.

Stemming from above, we can now formulate the design of the quantization precoded vector ${\bf u}_k$ for a given user $k$, and a given \textcolor{black}{FM} $\tightoverset{\rightarrow}{d'_k}$, as the following constrained convex optimization problem
\begin{alignat}{2} \label{eq:convex_problem}
	&\text{minimize}: \quad &&{\bf a}^T{\bf r}_k \nonumber\\ 
	&\text{subject to}: &&{\bf B}_k{\bf r}_k \preceq 0 \nonumber \\
	&  &&({\bf W}{\bf r}_k)^T({\bf W}{\bf r}_k)\le \frac{P_0}{ 2N_u P_g} \nonumber \\
	&  &&({\bf D}{\bf r}_k)^T({\bf D}{\bf r}_k)\le \alpha P'_0
\end{alignat}
where the corresponding vectors and matrices are 
\begin{align} \label{eq:optvec}
&{\bf r}_k=[{\bf u}'_k, -\gamma'_k] \nonumber \\
 &{\bf a}=[{\bf 0}_{1\times N_q}, 1]^T \nonumber \\
&{\bf B}_k=-[{\bf C}'_k {\bf G}_{tot,k}, {\bf 1}_{N_{tot}\times 1}] \nonumber \\
 &{\bf W}=[{\bf I}_{N_q}, {\bf 0}_{N_q\times 1}] \nonumber \\
&{\bf D}={\bf V}{\bf F}{\bf W}.
\end{align}
In above, ${\bf u}'_k$ and ${\bf C}'_k$ are the quantization precoded symbol vector and the codeword matrix, respectively, corresponding to \textcolor{black}{FM} $\tightoverset{\rightarrow}{d'_k}$, $N_{tot}=\mu_{RX}N_{block}$, $P_g=\frac{1}{T_u}\int_{-\infty}^{\infty}|g_{TX}(t)|^2dt$ is the normalized transmit filter energy, ${\bf F}$ is the $N-$point DFT matrix of size $N_q \times N$ with $(n,m)$'th element being $e^{-j\frac{2\pi nm}{N}}$ (see Appendix B), ${\bf V}=\text{diag}({\bf v})$, ${\bf v}=[{\bf 0}_{1\times p_1}, {\bf 1}_{1\times N-2p_1}, {\bf 0}_{1\times p_1}]^T$, while $p_1$ and $P'_0$ are given in Appendix B. The second constraint is due to the BS transmit power constraint and is elaborated in detail in Appendix B, see \eqref{eq:final_pow_cons}. The third constraint is due to the bandwidth constraint of the signal transmitted from each antenna branch and as elaborated in detail in Appendix B, the parameter $0<\alpha\ll1$ determines the amount of out-of-band emissions which is typically dictated by the emission constraints of the system. \textcolor{black}{In \eqref{eq:convex_problem}, the optimization variables are the $N_{block}\mu_{TX} \times 1$ quantization precoder output vector ${\bf u}_k$ and the magnitude of minimum unquantized received sample, $\gamma_k$, which are stacked into the vector ${\bf r}_k$ as shown in (16). Then for given FM, input data of block-size $N_{block}$, bandwidth constraint parameter $\alpha$, TX and RX filter coefficients, and TX and RX oversampling factors $\mu_{TX}$ and $\mu_{RX}$, the optimization problem in \eqref{eq:convex_problem} is solved for ${\bf r}_k$.} 

Furthermore, the overall optimization problem can be solved for each feasible $\tightoverset{\rightarrow}{d'_k}$ and the minimum out of the solutions is finally chosen. Note that ${\bf a}^T{\bf r}_k=-\gamma'_k$ and a positive $\gamma'_k$ can be found if and only if there exist a quantization precoded vector ${\bf u}'_k$ which yields quantized received vectors ${\bf z}_{RX,k}(i)=\text{sign}({\bf{s}}_{RX,k}(i))$ such that the \textcolor{black}{FM} $\tightoverset{\rightarrow}{d'_k}$ is satisfied for each $1\le i \le N_{block}$. Among all solutions of such ${\bf u}'_k$,  $\tightoverset{\rightarrow}{d'_k}$ and ${\bf C}'_k$, the ones that yield the largest positive $\gamma_k$ are then finally selected.

It is useful to note that optimizing the quantization precoder output samples with the constraint given in \eqref{eq:final_pow_cons} (Appendix B) implies $\text{trace}\left( \tilde{\bf S}^H \tilde{\bf S}\right)=\frac{P_0}{P_g}$ as given by \eqref{eq:pow_cons} if and only if $({\bf u}^{I}_k)^T({\bf u}^{Q}_k)=0$ and $\tilde{\bf A}^H\tilde{\bf A}= {\bf I}_{N_u}$ hold as strict equalities. However, in general, these may hold only approximatively since ${\bf u}^{I}_k$ and ${\bf u}^{Q}_k$ are designed independently while $\tilde{\bf A}^H\tilde{\bf A}= {\bf I}_{N_u}$  becomes more and more accurate with increasing number of antennas, i.e., holds asymptotically in the number of antennas. Hence, in order to comply with the transmit power constraint, first ${\bf u}^{I}_k$'s and ${\bf u}^{Q}_k$'s are designed via \eqref{eq:convex_problem} with the constraint derived in \eqref{eq:final_pow_cons}. Then, the initial quantization and channel precoded matrices are obtained from these vectors, denoted here as $\tilde{\bf U}'$ and $\tilde{\bf S}'=\tilde{\bf A}\tilde{\bf U}'$. The final precoded matrix is then obtained via scaling as $\tilde{\bf S}=c\tilde{\bf S}'$ where $c = \sqrt{\frac{P_0/P_g}{\text{trace}\left( (\tilde{\bf S}')^H (\tilde{\bf S}')\right)}}$, which implies $\text{trace}\left( \tilde{\bf S}^H \tilde{\bf S}\right)=\frac{P_0}{P_g}$.

The complete optimization-based framework for quantization precoder design can finally be formulated in an algorithmic form as follows:
\begin{center}
\fbox{
\parbox{0.95\linewidth}{{\bf{Algorithm:}} Design of Quantization Precoded Symbols
\begin{Algorithm}
	\item Set user index $k=1$. Go to step 2.
	\item If $k>N_u$, finish the procedure, else initialize $\gamma_k=0$, find all possible \textcolor{black}{FM}s  $\tightoverset{\rightarrow}{d_k}$. Mark all \textcolor{black}{FM}s as unused, go to step 3.
	\item If all \textcolor{black}{FM}s are used, set $k=k+1$ and go to step 2. Else pick one of the unused \textcolor{black}{FM}s  $\tightoverset{\rightarrow}{d'_k}$ and based on that initialize ${\bf C}'_k$, go to step 4.
	\item Solve \eqref{eq:convex_problem} for ${\bf r}_k$, if $\gamma'_k \le \gamma_k$, mark  $\tightoverset{\rightarrow}{d'_k}$ as used, go to step 3. If $\gamma'_k>\gamma_k$, set $\gamma'_k=\gamma_k$, ${\bf C}_k={\bf C}'_k$, $\tightoverset{\rightarrow}{d_k}=\tightoverset{\rightarrow}{d'_k}$ and ${\bf u}_k={\bf u}'_k$, mark $\tightoverset{\rightarrow}{d'_k}$ as used, go to step 3. 
\end{Algorithm}
}
}
\end{center}

\subsection{Discussion and Practical Aspects} \label{sec:OtherAspects}

The convex constrained optimization problem given in \eqref{eq:convex_problem} can be solved via available convex optimization tools such as CVX \cite{cvx}. The overall complexity of the algorithm can be found by multiplying the complexity of solving \eqref{eq:convex_problem} for a given \textcolor{black}{FM} with the number of \textcolor{black}{FM}s $N_{map}={R_{out} \choose R_{in}}R_{in}!$. As an example, 16-QAM modulation where per real-dimension $R_{in}=4$ and an RX oversampling factor of $\mu_{RX}=2$ yield $R_{out}=4$ and $N_{map}=24$. The parameter $R_{in}$  depends on the input modulation whereas $R_{out}=2^{\mu_{RX}}$ is a function of the receiver oversampling factor. The complexity of the algorithm rapidly increases with higher input modulation orders and/or oversampling factors. Also note that, per given \textcolor{black}{FM}, the dimension of the vector of unknowns, ${\bf r}_k$, is $\mu_{TX}N_{block}+1$. Hence, the overall complexity depends on $R_{in}$, $\mu_{TX}$, $\mu_{RX}$ and $N_{block}$.

\textcolor{black}{In general, to implement the proposed detection method, the scheduled terminals should know which \textcolor{black}{FM} was selected by the base station in the second step of the optimization. In this paper, we consider the following two practical methods to obtain such forward mapping information (FMI) in practice:} 
\begin{enumerate}[label=\roman*)]
	\item \textcolor{black}{Use only fixed and known mapping, and omit the second maximization step over different \textcolor{black}{FM}s, or}
	
	\item \textcolor{black}{Use a dedicated pilot sequence or control channel to communicate the optimized \textcolor{black}{FM}.} 
\end{enumerate}
\textcolor{black}{While one can devise alternative pilot-based schemes, one straight-forward method that we consider can be described as follows:}
\begin{itemize}
	\item \textcolor{black}{At the base station, a dedicated pilot sequence, specifically of the form $b_1, \cdots, b_{R_{in}}$ is appended as a prefix to the data part of each block. Here, the values $b_i$ are the symbol values of the used alphabet, per real domain, that is, $X_{in} = \{b_1, b_2, ..., b_{R_{in}}\}$ forms the symbol alphabet per real domain. This appended part is known in advance by the receivers, and deliberately covers all the different symbol levels.} 
	
	\item \textcolor{black}{At the UE receiver side, the \textcolor{black}{FM} can then be deduced from the pilot portion of the processing block. This is now feasible since the receivers know that the transmitter has appended the known sequence $b_1, \cdots, b_{R_{in}}$ prior to the quantization precoding. Since the known sequence covers all the different symbols levels, the full \textcolor{black}{FMI} can be deduced.}
	
	\item \textcolor{black}{Then, the corresponding backward mapping of the inferred \textcolor{black}{FM} is used for the detection of the actual data symbols in rest of the processing block.}
\end{itemize} 
\textcolor{black}{It is obvious that such pilot or control channel based delivery of the FM information is subject to the effects of noise, residual IUI and other possible sources of interference. We will thus address the achievable system performance under this method in the numerical evaluations in Section \ref{sec:Simulations}, where corresponding simulation examples are provided.}

As an additional practical aspect, we also wish to acknowledge that in order to form the matrix ${\bf G}_{tot,k}$ and hence to solve \eqref{eq:convex_problem}, the BS needs to strictly-speaking know the TX and RX filter responses, or more specifically, the corresponding sampled impulse responses. On one hand, knowing the essential TX filter response is close to trivial, as it is part of the BS's own processing chain. On the other hand, knowing the RX filter responses is not necessarily as straight-forward, unless for example the system specifications are defined such that certain type of RX filtering is required in all user devices. While this topic is partially outside the main scope of this article, we address the impact of imperfect RX filter knowledge in the numerical evaluations in Section \ref{sec:Simulations}, where corresponding simulation examples are provided. {\color{black}Similarly, since the analog RX filters are commonly of the infinite impulse response (IIR) form, while our system model derivations strictly-speaking assume FIR RX filtering, we will also study in Section \ref{sec:Simulations} the impact and the resulting system performance using IIR RX filters.}

Finally, for the fairness of the technical presentation and contents, it is to be acknowledged that while the latency of the detector processing in the receiver is very small, the latency of the overall precoding processing in the BS transmitter is directly proportional to the symbol block-length $N_{block}$. This is because the precoder optimization process and the corresponding resulting mapping from the data symbols to the sequence of quantization precoded samples operate on the input symbol block as a whole.

\section{Numerical Evaluations and Analysis} \label{sec:Simulations}

\subsection{Basic Simulation Assumptions}
In this section, we present comprehensive numerical evaluations, in terms of simulated SER curves, for the complete multi-user system illustrated in Fig. \ref{fig:MIMOsystem1bit_TRX} where the quantization precoder outputs are designed based on the algorithm described in Section \ref{sec:Design of Quantization Precoded Sequence}. In all the simulations, unless otherwise stated, the used data modulation is 16-QAM (4-ASK per real-dimension), TX and RX oversampling factors are $\mu_{TX}=\mu_{RX}=2$, the input block size is $N_{block}=50$, the DFT size is $N=2\mu_{TX}N_{block}=200$ and $\alpha=10^{-3}$. In general, both single-user and multi-user scenarios are evaluated, and also the number of antennas assumed to be adopted in the BS is varied. In general, the per user SER values are always averaged over 100 independent input symbol block and channel response realizations, while the shown SER values are then further averaged across the users.

\textcolor{black}{Without loss of generality, the receive pulse shaping filter in the basic simulations is a square-root raised cosine \textcolor{black}{(RRC)} filter with roll-off factor of $\epsilon_{RX}=0.22$ and length of 6 data symbol intervals, i.e., $N_{RX}=6$. As discussed in more details in Section II and Appendix A, the RX filter bandwidth is always equal to $W_{RX}=W_{TX}=(1+\epsilon_{TX})/T_{sym}$ where $\epsilon_{TX}$ is the roll-off factor for transmission bandwidth and considered to be 0.22 in all the simulation examples. The TX filter, on the other hand, is a regular raised-cosine whose length is 6 precoded sample durations, i.e., $N_{TX}=6$. For the case with $\mu_{TX}=1$, the roll-off factor for TX filter is $\epsilon_{TX,g}=\epsilon_{TX}=0.22$ such that $W_{TX,g}=W_{TX}$. For $\mu_{TX}>1$, the TX filter roll-off factor $\epsilon_{TX,g}$ is taken to be 0.1.} In general, the chosen filters reflect only one possible realistic setup, but are in no way any limiting or critical factor in the proposed overall waveform design and optimization framework. Under the above filtering characteristics, $W_{TX}=(1+0.22)/T_{sym}$ and hence, when combined with the assumed modulation, the spectral efficiency $R=\text{log}_2(4)/1.22=1.64$ bits/s/Hz per real dimension and per user. Thus, when spatially multiplexing, e.g., $N_u=5$ users, the total system spectral efficiency is already $2 \times 5 \times 1.64 = 16.4$ bits/s/Hz.

In the baseline simulations, it is assumed that the BS has perfect knowledge of the TX and RX sampled filter coefficients while the case of having some uncertainty in this knowledge is also experimented. \textcolor{black}{In all experiments, we assume that the entries of MIMO channel $\bf H$ are zero-mean unit variance i.i.d. circularly symmetric complex Gaussian variables.} In all the results, the CSI used in the channel precoder calculations is always estimated via orthogonal uplink pilots transmitted simultaneously by all user terminals. Denoting the number of uplink pilots transmitted by each user as $N_p$, then $N_u \times N_p $ orthogonal pilot matrix $\tilde{\bf X}_p$ is formed by stacking each user pilot data as row vectors which satisfies $\tilde{\bf X}^H_p\tilde{\bf X}^H_p={\bf I}_{N_u}$. The transmitted uplink pilots are then received at BS as
\begin{equation}\label{eq:uplink}
	\tilde{\bf Y}_p=\sqrt{\rho_u}\tilde{\bf H}^T_p\tilde{\bf X}_p+\tilde{\bf{N}}_p=\tilde{\bf{S}}_p+\tilde{\bf{N}}_p
\end{equation} 
where $\tilde{\bf{N}}_p$ denotes the receiver noise matrix whose entries are zero-mean i.i.d. circularly-symmetric complex Gaussian variables with variance $1$, and \textcolor{black}{$\rho_u$ is the uplink SNR which is set to 15 dB's}. \textcolor{black}{We note that in the uplink reception at the BS, we assume regular high-precision ADCs and thus the quantization effects are ignored unlike in the downlink transmission phase. Based on the model in \eqref{eq:uplink}, the channel estimate $\hat{\bf H}$ can be obtained via, e.g., the minimum-mean squared estimation (MMSE) \cite{ngo_massive_2013} while the actual channel precoder is then calculated by \eqref{eq:MRTZF} using the estimated channel $\hat{\bf H}$. We adopt MRT precoding in the single-user case whereas in the multi-user scenarios ZF precoder is used.}

The transmitted SNR during the downlink phase is defined as 
\begin{equation}
\text{SNR}=\frac{P_0}{E\left[\tilde{\boldsymbol{\eta}}_{filt,k}^H \tilde{\boldsymbol{\eta}}_{filt,k}\right]} 
\end{equation}
where, without loss of generality, the total BS transmit power is set to $P_0=1$. Notice that this definition essentially relates the total transmit power and the noise power of an individual receiver, hence in a multiuser setting, the transmit SNR from an individual user stream perspective is actually smaller.

\textcolor{black}{In general, there exist inter-block effects when time-domain filtering or pulse-shaping is adopted in a block-wise manner. If the length of the overall filter spans $L$ symbol intervals, then $L-1$ symbols in each block are subject to such inter-block interference. However, due to the damping nature of the impulse response, among these $L-1$ symbols, only 1-3 experience signiﬁcant interference whereas for others, the receiver thermal noise already tends to dominate. In our performance evaluations, we will thus deliberately discard 1-3 symbols (depending on the used filter lengths) when calculating the SER. Note that, since the used block-length is much higher than the number of discarded symbols, this can be seen as a very minor issue, in particular when the block-lengths of practical systems can easily be hundreds or thousands of data symbols. We will also evaluate the system performance under IIR type of RX filters as well as in such scenarios where the forward mapping information is communicated over a pilot channel from the BS to the UEs.}

\textcolor{black}{For comparison purposes, we also consider two benchmark methods, namely $\infty$-bit and $2$-bit receivers where only symbol-rate linear spatial channel precoding (MRT or ZF) is adopted. The former one refers to the ideal case where there is no quantization at the receiver. The latter one refers to the case where the received signals are uniformly quantized with $2$-bits/sample resolution over the dynamic range of the ADC, i.e., there are 4 quantization levels. In any practical receiver chain with multi-bit quantization, an automatic-gain-control (AGC) unit is needed that scales the incoming noisy signal to fit to the dynamic range of the ADC.  Typically, there is a certain offset or safety margin between the voltage range of the amplitude-controlled signal at the ADC input and the peak-to-peak support of the ADC circuitry. In the following evaluations, we assume that this offset is $10\%$. Moreover, the effective gain or scaling of the quantized useful signal component needs to be also acquired in order to perform the detection based on euclidean distance against the given multilevel reference constellation. For both $\infty$-bit and $2$-bit reference receivers, we assume that this effective gain is known up to a mean-squared estimation error of $20$ dB.}

Finally, even though some approximations were made in deriving the basic signal models for the waveform optimization purposes, no such approximations are made in the numerical evaluations in calculating the received signal. This holds in all studied scenarios. 
\subsection{Obtained Numerical Results}
\begin{figure}
    	\centering
        	\includegraphics[width=\linewidth]{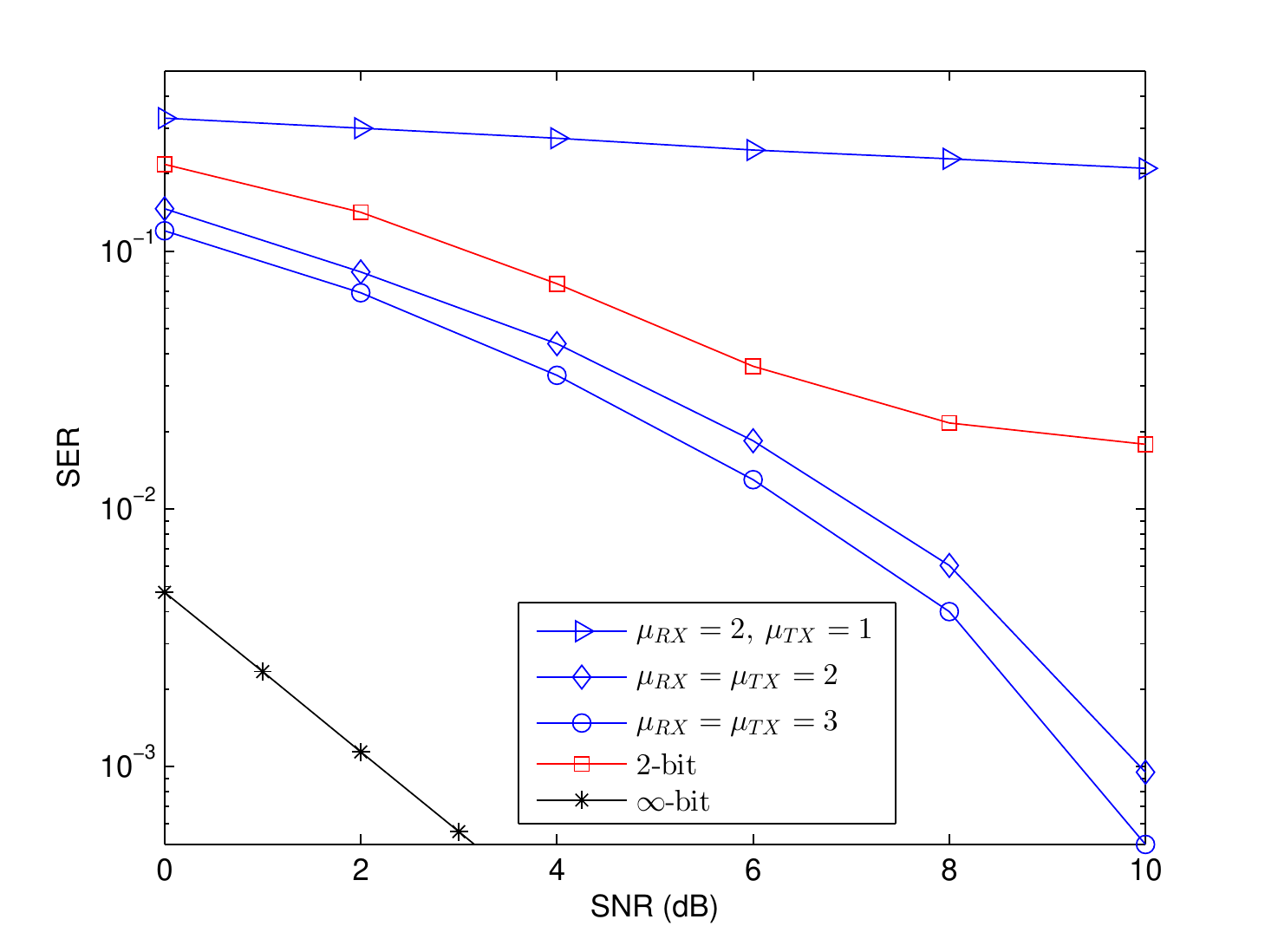} 
	\caption{SER vs. SNR in single-user scenario. Used modulation is 16-QAM  (4-ASK per real-dimension) and other fixed parameters are $N_t=50$, $N_{block}=50$, $N_{TX}=N_{RX}=6$, $\alpha=10^{-3}$. Spatial precoder is MRT and used $\mu_{TX}$ and $\mu_{RX}$ are indicated in the figure. \textcolor{black}{For reference, the SER curves of 2-bit and $\infty$-bit receivers with ordinary symbol-rate spatial precoding are also shown.}} 
	\label{fig:SERvsSNR_single_user}
\end{figure}

In Fig. \ref{fig:SERvsSNR_single_user}, the obtained SER characteristics vs. downlink SNR for different values of $\mu_{TX}$ and \textcolor{black}{$\mu_{RX}$} are illustrated, in the case of single-user transmission with $N_t=50$. It is observed that the SER is superior in the case where quantization precoding is performed with oversampled rate matched to the RX oversampling factor i.e., $\mu_{TX}=\mu_{RX}=2$ compared to the case with quantization precoding at data symbol rate, i.e., $\mu_{TX}=1$. \textcolor{black}{Moreover, when the oversampling rate is increased to $\mu_{RX}=\mu_{TX}=3$, even better SER performance is achieved due to increased Hamming distance between the valid codewords. However, we also observe that increasing $\mu_{RX}$ increases also the number of constraints of the optimization problem in \eqref{eq:convex_problem} and thus the obtained $\gamma$ values can decrease compared to the case with $\mu_{TX}=\mu_{RX}=2$. Therefore, despite the increased Hamming distance, the overall improvement in SER is rather modest. On the other hand, the computational complexity increases significantly with increasing $\mu$. Already in the case of $\mu_{TX}=\mu_{RX}=3$ and 4-ASK per real dimension, there is 5-times increase in the search space of forward mappings, compared to the case of $\mu_{TX}=\mu_{RX}=2$ whereas the dimensionality of the optimization problem for a given forward mapping increases as well. We thus conclude that $\mu_{TX}=\mu_{RX}=2$ provides a good performance vs. complexity trade-off and hence will be the default setting in all the following evaluations. It is also seen from Fig. \ref{fig:SERvsSNR_single_user} that the proposed techniques for 1-bit receivers outperform the benchmark 2-bit receiver. The performance of $\infty$-bit receiver is also shown for reference.}       

\begin{figure}
    	\centering
        	\includegraphics[width=\linewidth]{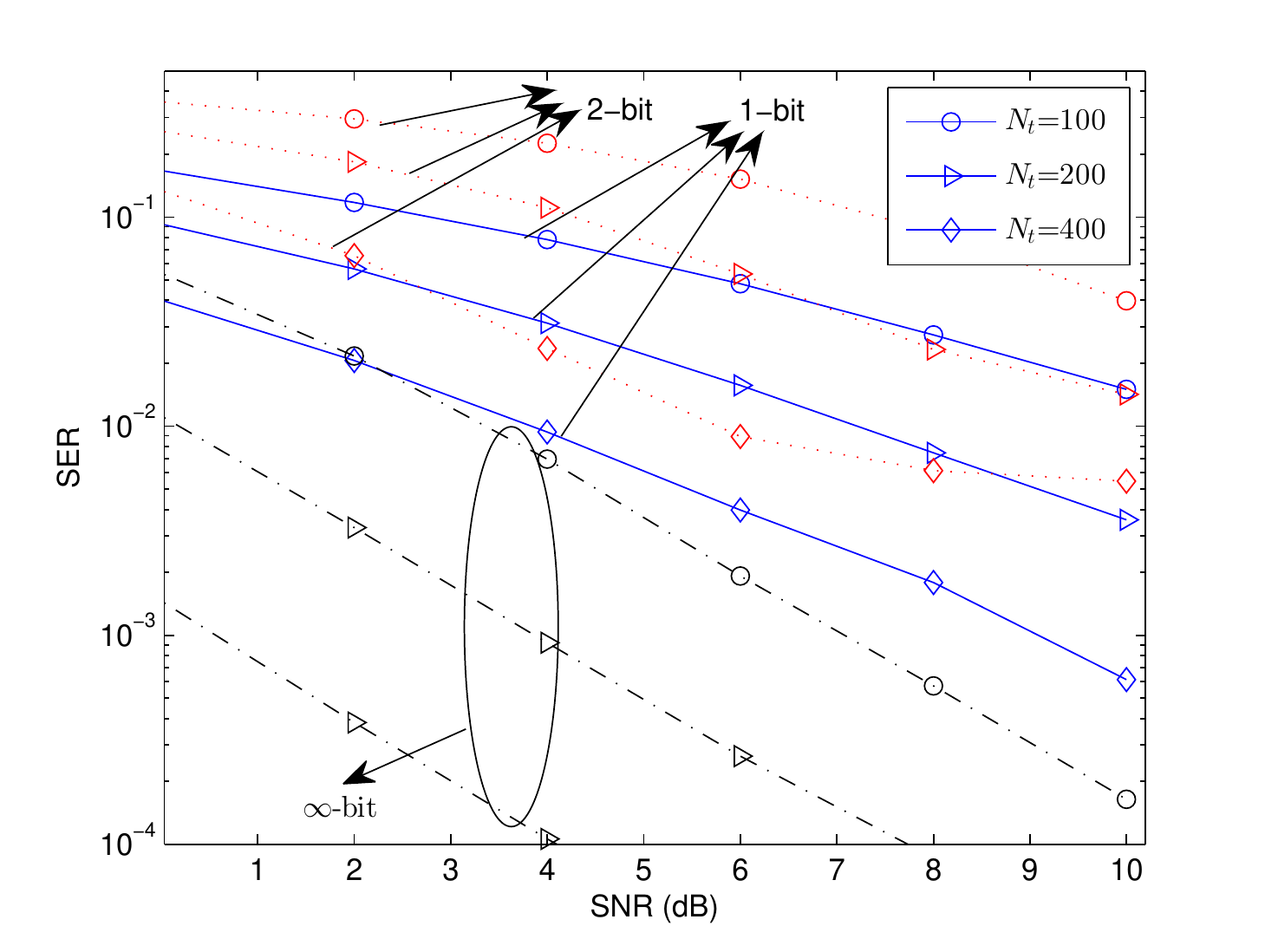} 
	\caption{SER vs. SNR in multi-user scenario with $N_u=5$ users. Used modulation is 16-QAM  (4-ASK per real-dimension) and other fixed parameters are $\mu_{TX}=\mu_{RX}=2$, $N_{block}=50$, $N_{TX}=N_{RX}=6$, $\alpha=10^{-3}$. Spatial precoder is ZF and used number of antennas are indicated in the figure. \textcolor{black}{For reference, the SER curves of 2-bit and $\infty$-bit receivers with ordinary symbol-rate spatial precoding are also shown.}} 
	\label{fig:SERvsSNR_multi_Nt}
\end{figure}

In Fig. \ref{fig:SERvsSNR_multi_Nt}, the obtained SER performance is illustrated in the multi-user case with $N_u=5$ users and \textcolor{black}{for indicated values of TX antennas, namely $N_t=100$, 200 and 400. As expected, the SER improves with increasing number of antennas since higher beamforming gains and better IUI suppression is achieved.} \textcolor{black}{However, already with $N_t=100$ a good SER performance of $10^{-2}$ is achieved under SNR$=10$ dB. It is also observed that for all choices of $N_t$, the proposed waveform optimization for 1-bit receivers outperforms the reference $2$-bit receiver cases. The performance of the reference $\infty$-bit receiver is again also shown.}

\begin{figure}
	\centering
	\includegraphics[width=\linewidth]{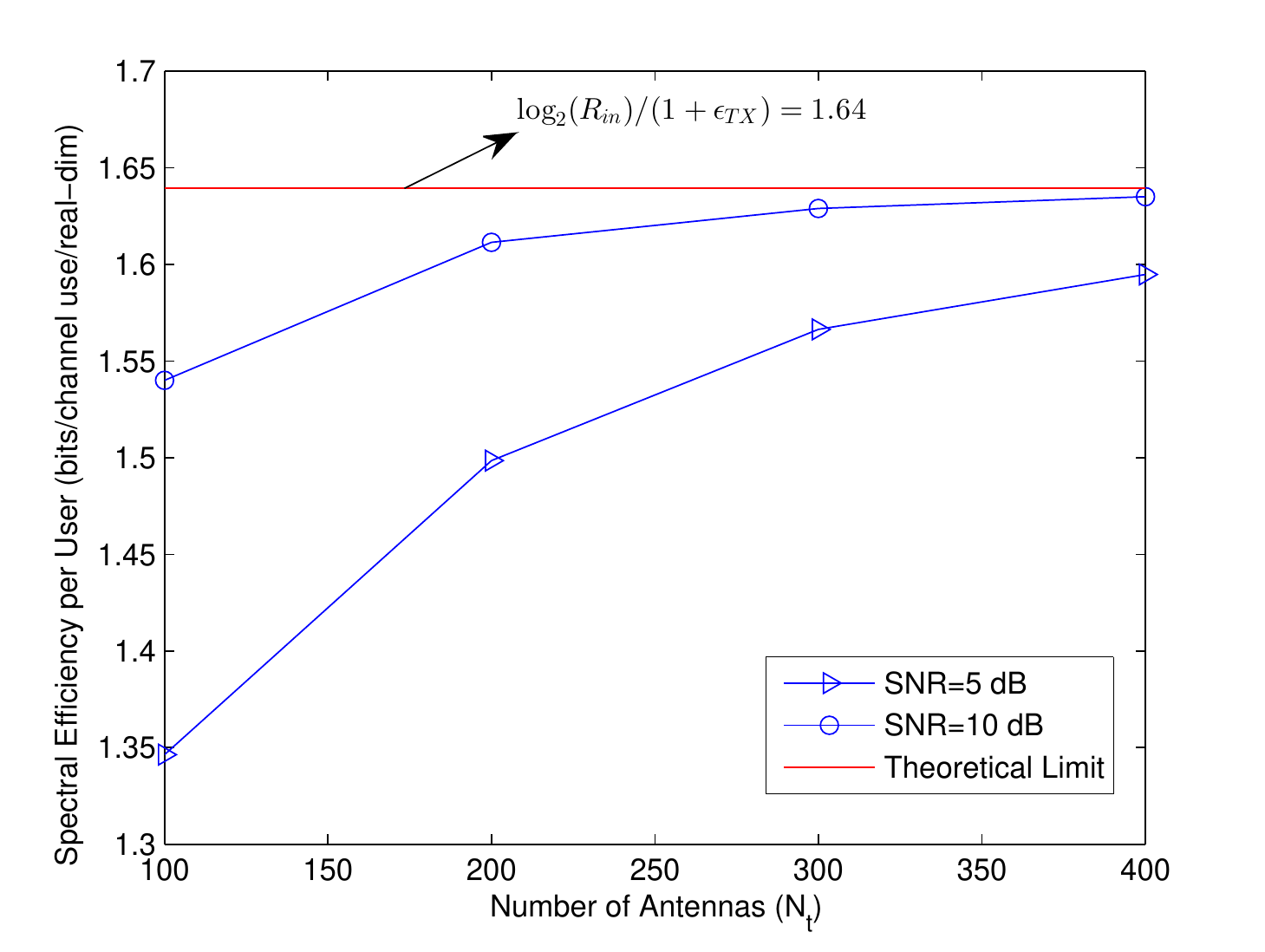} 
	\caption{\textcolor{black}{Spectral efficiency vs. $N_t$ in multi-user scenario with $N_u=5$ users. Used modulation is 16-QAM  (4-ASK per real-dimension) and other fixed parameters are $\mu_{TX}=\mu_{RX}=2$, $N_{block}=50$, $N_{TX}=N_{RX}=6$, $\alpha=10^{-3}$. Spatial precoder is ZF and used SNRs are indicated in the figure.}} 
	\label{fig:Mutual_Inf}
\end{figure}

\textcolor{black}{Next, the achievable mutual information is evaluated based on the simulated marginal and transition probabilities of the transmitted and received symbols and the corresponding pair-wise error scenarios, respectively. The obtained empirical values are averaged over different users and the obtained per user spectral efficiencies are plotted in Fig. \ref{fig:Mutual_Inf} for different numbers of antennas and indicated SNR values. When SNR$=$10 dB and $N_t=100$, the proposed waveform optimization achieves already 1.54 bits/channel use/real-dim which is 94$\%$ of the theoretical maximum spectral efficiency of $\text{log}_2(4)/(1.22)=1.64$ while for larger numbers of the TX antennas, the realized mutual information is even closer to the theoretical upper limit. For lower SNR of 5 dB, the mutual information levels are slightly lower but still, e.g. with $N_t=100$ antennas, the realized mutual information of 1.35 is approximately 82$\%$ of the theoretical maximum. This demonstrates the good efficiency of the proposed methods.}

\begin{figure}
    	\centering
        	\includegraphics[width=\linewidth]{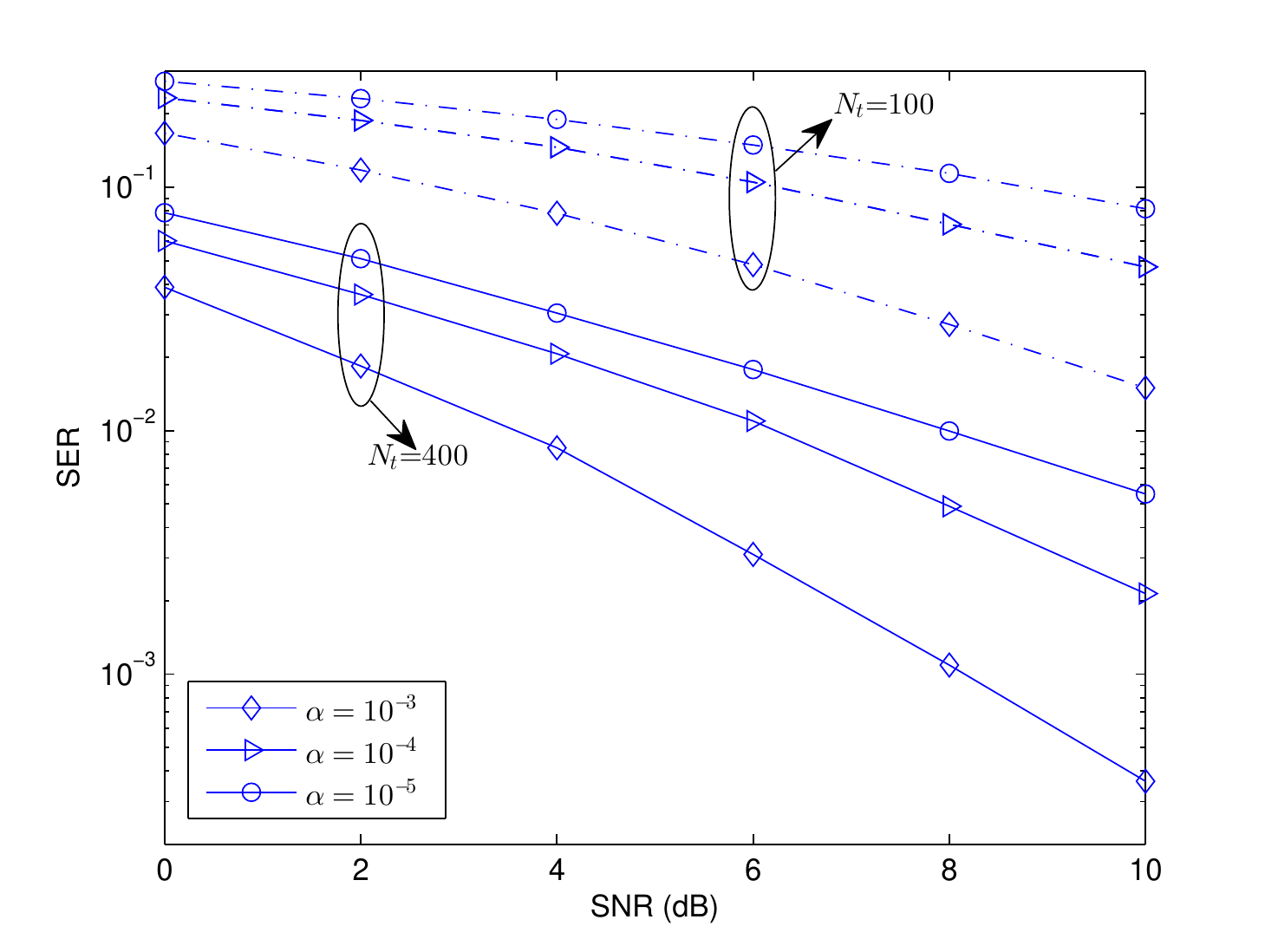} 
	\caption{SER vs. SNR in multi-user scenario with $N_u=5$ users. Used modulation is 16-QAM  (4-ASK per real-dimension) and other fixed parameters are $\mu_{TX}=\mu_{RX}=2$, $N_{block}=50$, $N_{TX}=N_{RX}=6$. Spatial precoder is ZF, while used $N_t$ and $\alpha$ values are indicated in the figure.} 
	\label{fig:SERvsSNR_alpha}
\end{figure}

Next, in Fig. \ref{fig:SERvsSNR_alpha}, the obtained SER performance is illustrated with respect to the factor $\alpha$ that determines the out-of-band emission power being regulated by unwanted emission limits. \textcolor{black}{In this example, the number of users is $N_u=5$ whereas two different settings are used for the number of antennas, namely $N_t=100$ and $N_t=400$. For each setting of $N_t$, SER performance is experimented with three different values of $\alpha$, namely $10^{-3}$, $10^{-4}$ and $10^{-5}$. It is seen that the SER performance degrades with decreasing $\alpha$ since the out-of-band emission constraint in \eqref{eq:convex_problem} gets more strict.} It can be concluded that for more strict spectral mask constraints requiring smaller values of $\alpha$, more antennas are also needed to achieve a target SER value.

\begin{figure}
	\centering
	\includegraphics[width=\linewidth]{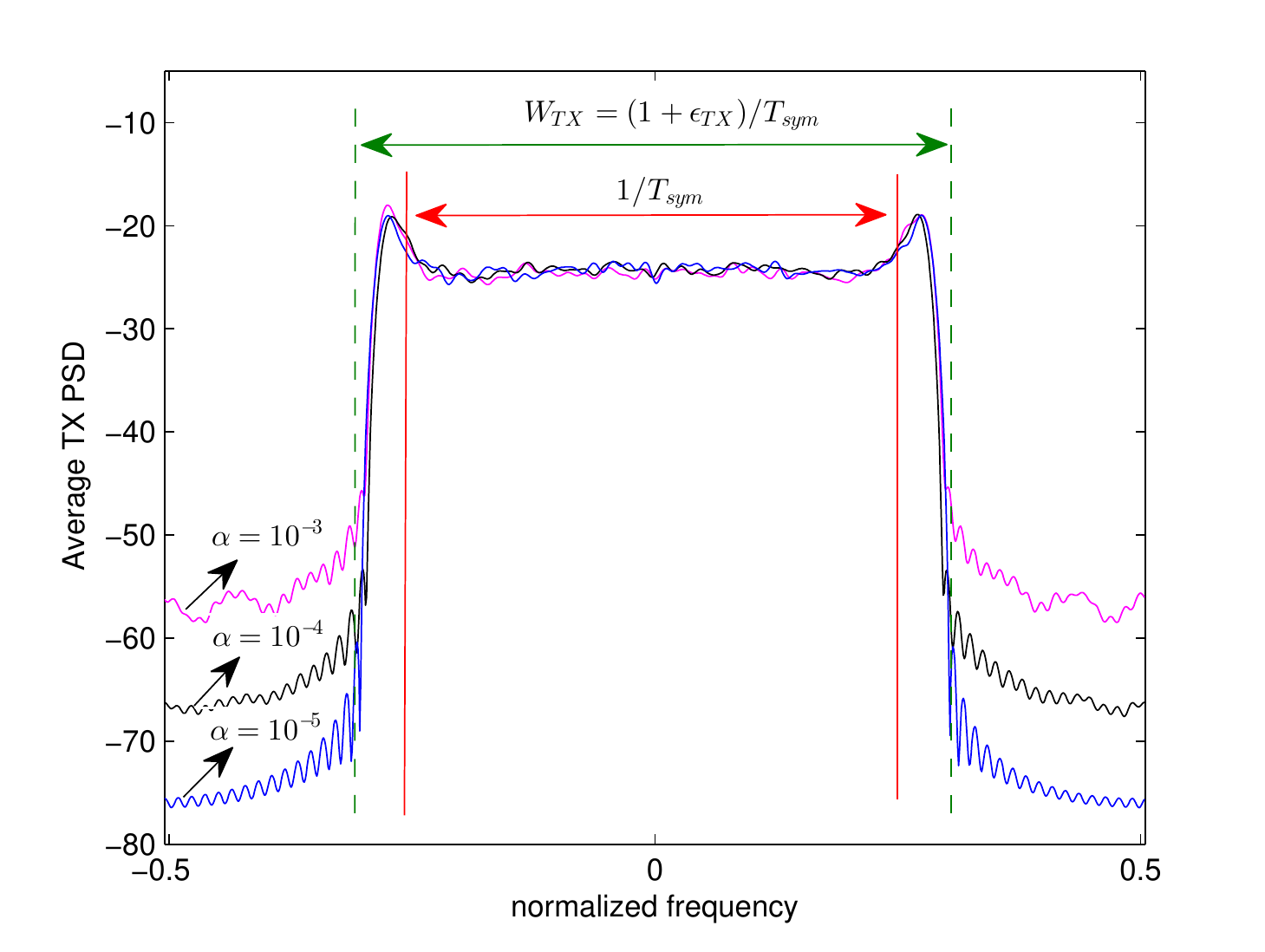} 
	\caption{\textcolor{black}{Power spectral density of the transmitted signal for indicated values of $\alpha$ when $\mu_{TX}=2$.}}
	\label{fig:PSD}
\end{figure}

\textcolor{black}{Next, in order to demonstrate the impact of the bandwidth constraint parameter $\alpha$ on the TX waveform spectral characteristics, the power spectral density (PSD) of the transmitted signal is plotted in Fig. \ref{fig:PSD} for $N_u=5$, $N_t=100$ and indicated values of $\alpha$ when TX oversampling factor is $\mu_{TX}=2$. The used number of FFT-points for evaluating the spectrum is $N_{fft}=1024$ which is about 10-times higher than the individual (oversampled) block-size and hence provides good frequency resolution. In obtaining the curves, first, the PSD of a particular transmit block at a particular TX antenna is calculated and then averaging is performed over different transmit blocks and antennas. Note that for $\mu_{TX}=2$, the transmitted signal should ideally be confined to the band whose borders are shown with the green dash-dotted line. As indicative example values, when $\alpha=10^{-3}$, the out-of-band suppression is in the order of $34$ dB whereas it is already around $44$ dB for $\alpha=10^{-4}$. Then, for a more strict value of $\alpha=10^{-5}$, the out-of-band suppression is already approximately $54$ dBs. In practice, to achieve the lowest SER, $\alpha$ can be set to highest possibe value that still satisfies the system spectral mask constraint.}

\begin{figure}
    	\centering
        	\includegraphics[width=\linewidth]{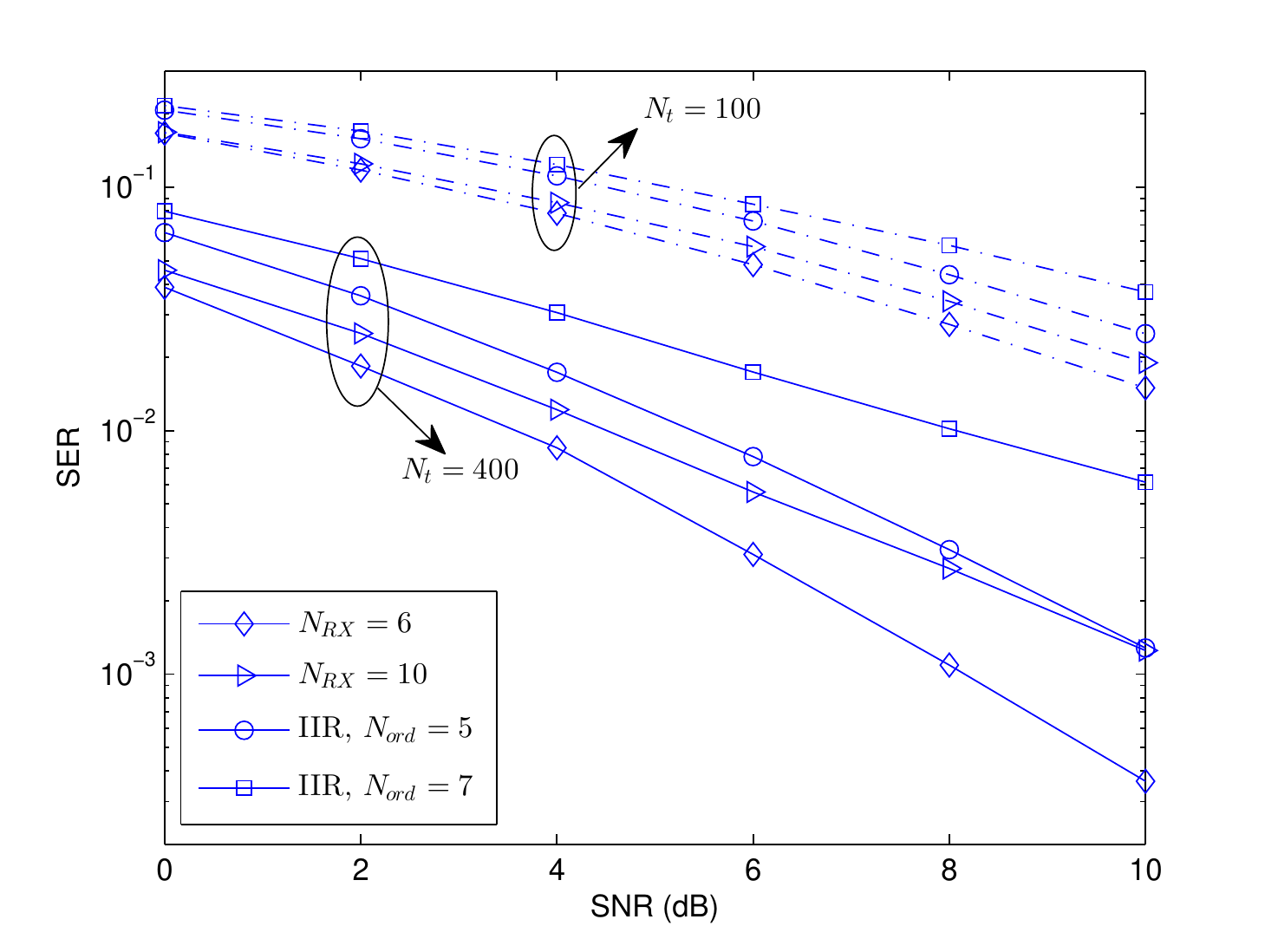} 
	\caption{SER vs. SNR in multi-user scenario with $N_u=5$ users. Used modulation is 16-QAM  (4-ASK per real-dimension) and other fixed parameters are $\mu_{TX}=\mu_{RX}=2$, $N_{block}=50$, $N_{TX}=6$, $\alpha=10^{-3}$. Spatial precoder is ZF, used values of $N_t$ and receiver filter parameters are indicated in the figure.}  
	\label{fig:SERvsSNR_gRX}
\end{figure}

In Fig. \ref{fig:SERvsSNR_gRX}, the obtained SER performance is illustrated with respect to \textcolor{black}{indicated receive filters. For RRC filters, we observe that the SER is only mildly degraded when $N_{RX}$ is increased from 6 to 10. From Fig. \ref{fig:SERvsSNR_multi_Nt}, one can conclude that for fixed $N_{RX}$ and $N_u$, increasing the number of antennas improves the SER. Hence, we can state that for longer receive filter lengths, and hence longer total effective ISI response, more antennas are needed to achieve a target SER value. Next, since analog RX filters used in practice are commonly of the IIR type, the achievable performance with such RX filtering is explored as well. As a practical example, we consider IIR Butterworth filters of order $N_{ord}=5$ and 7 whereas the cut-off frequency is one quarter of the sample rate (RX oversampling factor is here 2). While the actual RX filtering is implemented with the true IIR Butterworth filter, a truncated FIR approximation of length 8 symbol durations is adopted in the precoder optimization phase. This is because the optimization formulation in \eqref{eq:convex_problem} assumes FIR filtering. The obtained results in Fig. \ref{fig:SERvsSNR_gRX} demonstrate that the proposed waveform design and optimization procedure can be successfully applied even in the cases where the RX filter is an IIR filter whose impulse response is strictly-speaking infinitely long. To this end, it is interesting to note that the overall optimization problem formulated in  \eqref{eq:convex_problem} depends on the chosen TX and RX filter coefficients via the matrix $\mathbf{B}_k$. In other words, different choices of TX and RX filters lead to different matrix structures for $\mathbf{B}_k$ which, in turn, will affect the optimization problem and thus the optimized values of the quantization precoder output and thereon the achievable SER. Thus, one can seek to further optimize the system performance by varying the used TX and RX filters, executing the optimization for each given pair of TX and RX filters, and finally choosing such TX and RX filters which yield the best system performance. Such filter optimization is an interesting topic for our future research.}

\begin{figure}
	\centering
	\includegraphics[width=\linewidth]{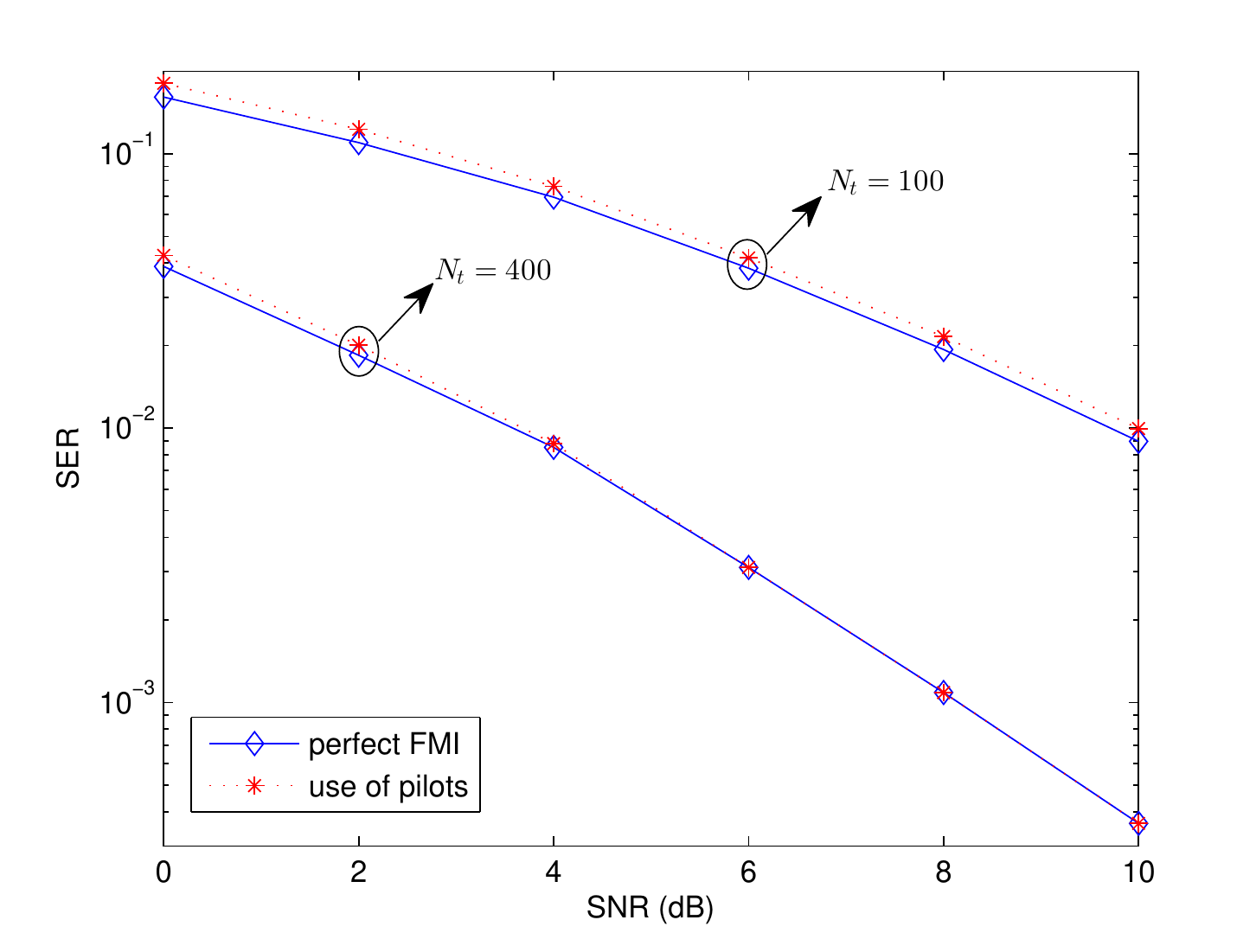} 
	\caption{\textcolor{black}{SER vs. SNR in multi-user scenario with $N_u=5$ users under different FM scenarios. Used modulation is 16-QAM  (4-ASK per real-dimension) and other fixed parameters are $\mu_{TX}=\mu_{RX}=2$, $N_{block}=50$, $N_{TX}=N_{RX}=6$, $\alpha=10^{-3}$. Spatial precoder is ZF and used $N_t$ values are indicated in the figure.}}  
	\label{fig:SERvsSNR_FMI}
\end{figure}

\textcolor{black}{Next, we explore the achievable system performance under the practical pilot based method to reveal the used FM to the intended receivers. In Fig. \ref{fig:SERvsSNR_FMI}, the corresponding SER curves are plotted for $N_u=5$, $N_t=100, 400$, and for two cases of (i) perfect FMI (shown for reference) and (ii) estimated FMI obtained using the downlink pilots. In the pilot based scheme, only 4 symbols are pilots since with 16-QAM, $R_{in}=4$. As can be seen from the figure, with only a small pilot overhead in each block, the proposed pilot scheme achieves practically identical performance compared to the perfect FMI case.}

In all previous evaluations, the base-station was assumed to have perfect knowledge of the RX filter sampled impulse responses, whereas next we study the impact of imperfect knowledge of it, i.e., of $g_{RX,k}(n)$. We consider a model where the coefficients known to BS are of the form $\hat{g}_{RX,k}(n)={g}_{RX,k}(n)+\eta_{g,k}(n)$ for $0\le n \le \mu_{RX}N_{RX}$ where $\eta_{g,k}(n)$ is real-valued zero-mean i.i.d. Gaussian noise with variance $\sigma^2_g$. In vector notation, we can write $\hat{\bf g}_{RX,k}={\bf g}_{RX,k}+\boldsymbol{\eta}_{g,k}$ where ${\bf g}_{RX,k}=[{g}_{RX,k}(0), \cdots, {g}_{RX,k}(\mu_{RX}N_{RX})]^T$ and $\boldsymbol{\eta}_{g,k}=[\eta_{g,k}(0), \cdots, \eta_{g,k}(\mu_{RX}N_{RX})]^T$. Then, we define the so-called filter-to-noise ratio as $\text{FNR}=\frac{{\bf g}^T_{RX,k}{\bf g}_{RX,k}}{E\left[|{\eta}_{g,k}(n)|^2\right]}$ to quantify the level of assumed modeling uncertainty. In order to see clearly the effect of such imperfect knowledge of the RX filter coefficients, we assume perfect CSI and zero-noise during downlink transmission, i.e., $\text{SNR}=\infty$. 

\begin{figure}
    	\centering
        	\includegraphics[width=\linewidth]{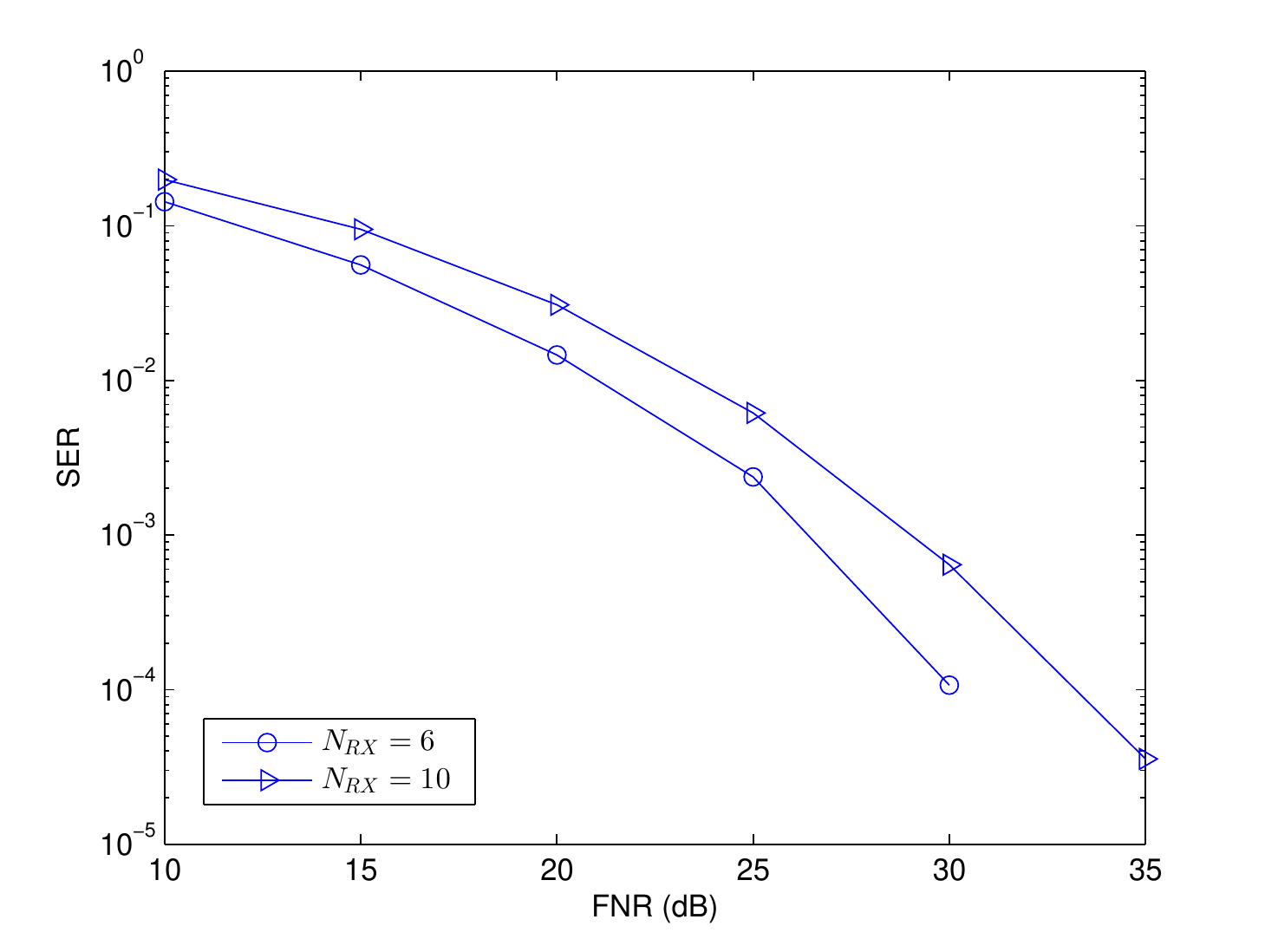} 
	\caption{SER vs. FNR in multi-user scenario with $N_u=5$. Used modulation is 16-QAM  (4-ASK per real-dimension) and other fixed parameters are $N_t=100$, $\mu_{TX}=\mu_{RX}=2$, $N_{block}=50$, $N_{TX}=6$, $\alpha=10^{-3}$, SNR$=\infty$. Spatial precoder is ZF and used $N_{RX}$ values are indicated in the figure.}  
	\label{fig:SERvsFNR}
\end{figure}

\textcolor{black}{In Fig. \ref{fig:SERvsFNR}, the SER performance is plotted against the FNR for indicated receive filter lengths when the number of antennas and the number of users are set to $N_t=100$ and $N_u=5$, respectively. We observe that the SER performance depends on the RX filter length, and in general, the required FNR in order to have negligible influence on the SER performance depends on the RX filter length. For instance, $\text{FNR}=30$ dB yields $\text{SER}\approx 10^{-4}$ for $N_{RX}=6$ whereas similar SER is reached approximately for $\text{FNR}=33$ dB when $N_{RX}=10$.}       

\section{Conclusions}  \label{sec:Conclusion}
In this article, we considered the problem of multiuser-multiantenna waveform design and optimization for massive MIMO downlink with highly simple 1-bit oversampled receivers, incorporating a general target of achieving spectral efficiencies higher than 1 bits/s/Hz per user and per real-dimension. A novel two-stage spatio-temporal precoder solution was proposed for the base-station transmitter, complemented by low-complexity Hamming-distance based symbol-detectors at terminals. The provided numerical examples show that good link and system performance can be achieved when the spatio-temporal precoding is done at oversampled rate, matched to the RX oversampling factor. In general, the obtained results indicate that the proposed spatio-temporal precoding and waveform optimization scheme, together with the low-complexity detector functionality, can enable efficient and reliable multiplexing of many low-cost low-complexity 1-bit receivers well suited for the IoT use cases in the emerging massive MIMO based 5G and beyond networks. In general, the waveform optimization developments carried out in this article are specifically focusing on narrowband channels with single-carrier modulation. Generalizing the spatio-temporal precoding and waveform solutions for wideband channels and/or multi-carrier radio links form interesting and important topics for our future work.

\appendices 
\section{Derivation of Received Signal Model} \label{app:TRX_process}

Here, we provide a detailed step-by-step derivation of the received signal model prior to 1-bit quantization, resulting effectively from the TX filtering of the precoded transmit samples, fading multi-user MIMO channel, RX filtering and finally sampling.

\begin{figure}
    	\centering
        	\includegraphics[width=\linewidth]{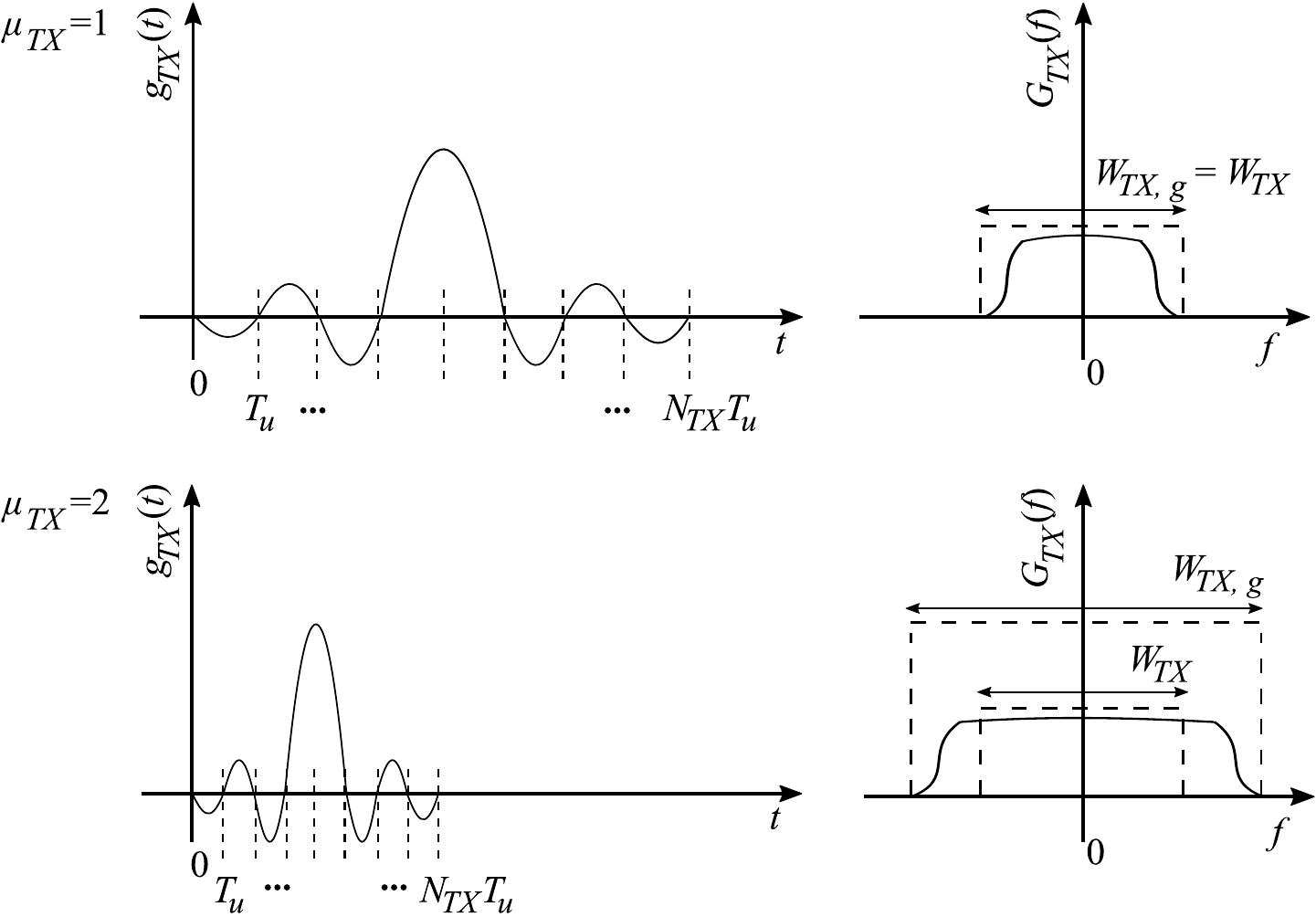} 
	\caption{Example impulse and frequency responses of the transmit filter, $g_{TX}(t)$ and $G_{TX}(f)$, respectively, when precoded samples are at symbol-rate (upper) and twice the symbol-rate (lower), i.e., $\mu_{TX}=1$ and $\mu_{TX}=2$. The TX filter bandwidth $W_{TX,g}$ and the desired transmission bandwidth $W_{TX}$ are illustrated for both cases when symbol period is fixed to $T_{sym}$ and precoded symbol period at the output of TX precoder is $T_u=\frac{T_{sym}}{\mu_{TX}}$.}
	\label{fig:TX_filter}
\end{figure}

On the TX side, the precoded samples at the output of the TX precoder are first converted to continuous waveforms at the $I$ and $Q$ branches of each antenna unit via digital-to-analog converters (DACs) and TX filters. For simplicity, we assume that the TX filters are identical in all antenna chains. The transmit filter, denoted by $g_{TX}(t)$, is assumed to be a low-pass filter band-limited to a bandwidth \textcolor{black}{$W_{TX,g}=\frac{1+\epsilon_{TX,g}}{T_u}=\frac{(1+\epsilon_{TX,g})\mu_{TX}}{T_{sym}}$} where \textcolor{black}{$0\le \epsilon_{TX,g} \le 1$} is the roll-off factor. The impulse response of the TX filter $g_{TX}(t)$ is assumed to be zero outside of the interval $0\le t\le N_{TX}T_{u}$ where $N_{TX}\in \mathbb{Z}^+$ is a sufficiently large integer that satisfies the filter bandwidth requirement. For fixed $T_{sym}$, the TX filter duration and bandwidth depend on the precoded symbol period $T_u$, i.e., duration is shorter and bandwidth is larger when precoder output is at higher rate, as illustrated in Fig. \ref{fig:TX_filter}. 
Note that when $\mu_{TX}>1$, the TX filter has larger bandwidth $W_{TX,g}$ than the desired transmission bandwidth $W_{TX}$. However, as we elaborate in details in Section \ref{sec:Design of Quantization Precoded Sequence} and Appendix \ref{app:constraints}, the design and optimization of the TX precoder outputs includes an appropriate bandwidth constraints, in the form of controlled ISI, such that the continuous-time signal formed by TX-filtering the precoder output samples at each antenna branch is bandlimited to $W_{TX}$. \textcolor{black}{In this respect, the overall transmission scheme considered and developed in this work is inherently different from classical spread spectrum techniques, where the adoption of higher-rate transmit samples or chips induce expansion of the transmitted signal bandwidth.} 

Continuing then with the signal model, the transmitted continuous-time signal at the $l$'th antenna branch reads
\begin{equation} \label{eq:TX_signal_l}
	\tilde{s}_{TX,l}(t)=\sum_{q=0}^{N_q-1} \tilde{s}_l(q)g_{TX}(t-qT_{u})=s^I_{TX,l}(t)+js^Q_{TX,l}(t) 
\end{equation}
where $s^{\theta}_{TX,l}(t)=\sum_{q=0}^{N_q-1} s^{\theta}_l(q)g_{TX}(t-qT_{u})$. The total transmitted continuous time signal, at time $t$ at the output of the transmitter antenna array with $N_t$ antenna units, can then be represented with the vector-valued waveform $\tilde{\bf s}_{TX}(t)=[s_{TX,1}(t), \cdots, s_{TX,N_t}(t)]^T$.

It is next assumed that the number of TX antennas in the massive MIMO BS is sufficiently large to allow channel hardening such that $\tilde{\bf{H}}\tilde{\bf{A}} \approx \text{diag}\left(\beta_1, \cdots, \beta_{N_u}\right)$ is tight for large $N_t$, e.g. $N_t>50$ \cite{hochwald_multiple-antenna_2004} where $\beta_k>0$ refers to the real-valued beamforming gain for user $k$. Note that this assumption concerns only the MRT channel precoding, while the ZF precoding always fulfills this without any approximations (assuming perfect CSI). Under such assumption, the noiseless complex multi-user signal after fading channels, namely $\tilde{\bf s}_{CH}(t)=\tilde{\bf{H}}{\bf{s}}_{TX}(t)$, has such structure that the noiseless complex received signal of the $k$-th user reads 
\begin{align}\label{eq:TX_cont}
	{\tilde{s}}_{CH,k}(t)&\approx \beta_k \sum_{q=0}^{N_q-1} \tilde{u}_k(q)g_{TX}(t-qT_u) \nonumber \\
	&= {s}^I_{CH,k}(t)+ j{s}^Q_{CH,k}(t)
\end{align}
where ${s}^{\theta}_{CH,k}(t)= \beta_k \sum_{q=0}^{N_q-1} u^{\theta}_k(q)g_{TX}(t-qT_u)$ with the approximation sign being due to channel hardening. 

On the RX side, the $I$ and $Q$ components of ${\tilde{s}}_{CH,k}(t)$ are filtered via identical real-valued receive filters with impulse response $g_{RX,k}(t)$. The receive filters are assumed to be of low-pass form, essentially limiting the received signal double-sided bandwidth to the useful signal bandwidth $W_{TX}$ and suppressing all spectral components outside this band. The impulse response of the RX filter $g_{RX,k}(t)$ is assumed to be zero outside the interval of $0\le t\le N_{RX}T_{sym}$ where $N_{RX}$ is the receive filter length in input-symbol periods. The noisy complex signal after receive filtering can now be written as
\begin{align}\label{eq:sRX_t}
\tilde{y}_{RX,k}(t) &=(\tilde{s}_{CH,k}(t)+\tilde{\eta}_k(t)) \otimes g_{RX,k}(t)\nonumber \\
&=\tilde{s}_{RX,k}(t)+\tilde{\eta}_{filt,k}(t)
\end{align}
where $\tilde{s}_{RX,k}(t)=\tilde{s}_{CH,k}(t) \otimes g_{RX,k}(t)$ and  $\tilde{\eta}_{filt,k}(t)=\tilde{\eta}_k(t) \otimes g_{RX,k}(t)$ are the filtered signal and noise terms, respectively. The additive noise before filtering is given as $\tilde{\eta}_k(t)=\eta^I_k(t)+j\eta^Q_k(t)$ where $\eta^I_k(t)$ and $\eta^Q_k(t)$  are i.i.d. zero-mean real-valued white Gaussian noise signals on $I$ and $Q$ branches with spectral density $\frac{N_0}{2}$, respectively. 

The analog receive filtering is followed by the 1-bit ADCs in the $I$ and $Q$  branches which incorporate sampling and 1-bit quantization. The sampled signal prior to quantization is denoted as
\begin{align}\label{eq:sRX_m}
\tilde{y}_{RX,k}(n) &\doteq \tilde{y}_{RX,k}\left(nT_{s}\right)={y}^I_{RX,k}(n)+j{y}^Q_{RX,k}(n)
\end{align}
where $T_s=\frac{T_{sym}}{\mu_{RX}}$ is the sampling period with $\mu_{RX}$ being the oversampling factor at RX and ${y}^{\theta}_{RX,k}(n)={s}^{\theta}_{RX,k}(n)+{\eta}^{\theta}_{filt,k}(n)$ with ${s}^{\theta}_{RX,k}(n) \doteq {s}^{\theta}_{RX,k}(nT_s)$, ${\eta}^{\theta}_k(n)  \doteq {\eta}^{\theta}_k(nT_s)$. Stemming from the sampling theorem, the useful signal term can be written as
\begin{align}
\tilde{s}_{RX,k}(n)&={s}^I_{RX,k}(n)+j{s}^Q_{RX,k}(n)=\tilde{s}_{CH,k}(n) \otimes g_{RX,k}(n) \nonumber \\
&=\sum_{m=0}^{L_{RX}}\tilde{s}_{CH,k}(n-m)g_{RX,k}(m)
\end{align}
 where  $L_{RX}=\mu_{RX}N_{RX}$ is the length of the sampled RX filter $g_{RX,k}(n)\doteq g_{RX,k}(nT_s)$ and $\tilde{s}_{CH,k}(n)\doteq \tilde{s}_{CH,k}(nT_s)$. Then, the useful signal term can be explicitly expressed as 
\begin{align} \label{eq:sRX_sample} 
	\tilde{s}_{RX,k}(n) &=\sum_{m=0}^{L_{RX}-1}\tilde{s}_{CH,k}(n-m)g_{RX,k}(m)  \nonumber \\
				& =\beta_k \sum_{q=0}^{N_q-1} \tilde{u}_k(q)g_{tot,k}(q,n) =\beta_k {\bf g}^T_{tot,k}(n)\tilde{\bf u}_k\nonumber \\
				&=\underbrace{\beta_k {\bf g}^T_{tot,k}(n){\bf u}^I_k}_ {{s}^I_{RX,k}(n)}+j\underbrace{\beta_k {\bf g}^T_{tot,k}(n){\bf u}^Q_k}_{{s}^Q_{RX,k}(n)}
\end{align}  
where $g_{tot,k}(q,n)$ denotes the total effective filter response. Furthermore, denoting $g_{TX,k}(n)\doteq g_{TX,k}(nT_s)$, we can write
\begin{align}
g_{tot,k}(q,n)=\sum_{m=0}^{L_{RX}-1}g_{TX}\left((n-m)-q\frac{\mu_{RX}}{\mu_{TX}}\right)g_{RX,k}(m).
\end{align}
Then, the total filter response vector reads ${\bf g}_{tot,k}(n)=[g_{tot,k}(0,n),\cdots, g_{tot,k}(N_{q}-1,n)]^T$. This concludes the derivation of the fundamental received signal model in \eqref{eq:sRX_m_text}-\eqref{eq:sRX_sample_theta}.

\section{Derivation of Power and Bandwidth Constraints}  \label{app:constraints}
Here, the transmission power and bandwidth constraints that are adopted in the convex optimization problem given by \eqref{eq:convex_problem}-\eqref{eq:optvec} are derived.

\subsection{Transmission Power Constraint} \label{app:power}
For the optimization purposes, we wish to establish a relation between the total BS transmit sum-power constraint, denoted by $P_0$, and the quantization precoded streams $\tilde{\bf u}_k$ for each user $k$. The normalized {\color{black}instantaneous} transmit sum-power {\color{black}for a composite block of $N_u \times N_{block}$ data symbols} can be expressed as 
\begin{align}\label{eq:pow_cons}
P_0&=\frac{1}{T_u}\sum_{l=1}^{N_t} \int_{-\infty}^{\infty} \mid \tilde{s}_{TX,l}(t) \mid^2 dt \nonumber \\
&=\sum_{l=1}^{N_t} \sum_{i=0}^{N_q-1}\sum_{j=0}^{N_q-1}\tilde{s}_l(i)\tilde{s}^*_l(j)R_g(i-j)  \nonumber \\
&=\sum_{l=1}^{N_t} P_g\tilde{\bf s}^H_l \tilde{\bf s}_l=P_g\text{trace}\left( \tilde{\bf S}^H \tilde{\bf S}\right) \nonumber \\
&=P_g\text{trace}\left((\tilde{\bf A}\tilde{\bf U})^H(\tilde{\bf A}\tilde{\bf U}))\right)
\end{align} 
where $R_g(i-j) =R_g(j-i)=\frac{1}{T_u}\int_{-\infty}^{\infty}g_{TX}(t-iT_u)g_{TX}(t-jT_u)dt$ is the normalized auto-correlation function of the transmit filter evaluated at lags $\mid i-j \mid T_u$. In the third line, we assume that the transmit filter is designed such that the normalized auto-correlation function satisfies $R_g(0)=P_g$ and $R_g(i)=0$ for $\mid i \mid> 0$. Note that this is true, e.g., for all square-root Nyquist pulses. Notice also that we have deliberately assumed the precoded samples, and hence the overall transmit waveform, deterministic, i.e., the above expression applies for arbitrary given precoded samples $\tilde{\bf u}_k, k=1,2,...,N_u$.

Now, we assume that the MIMO channel estimates are reasonably good and thus the channel hardening applies, implying $\tilde{\bf A}^H\tilde{\bf A}\approx {\bf I}_{N_u}$ given that the formal precoder scaling coefficients in \eqref{eq:MRTZF} are chosen as  $c_{MRT}=\sqrt{\frac{1}{N_t}}$ and $c_{ZF}=\sqrt{N_t}$. Then, the last line of \eqref{eq:pow_cons} can be well approximated as 
\begin{align} \label{eq:power_constraint}
P_0 \approx&P_g \text{trace}\left(\tilde{\bf U}^H\tilde{\bf U}\right)=P_g \sum_{k=1}^{N_u} \tilde{\bf u}^H_k\tilde{\bf u}_k
\end{align}
Now, in order to obtain a per-user constraint, we impose the condition that ${\bf u}^H_1{\bf u}_1=\cdots={\bf u}^H_{N_u}{\bf u}_{N_u}$ yielding the power constraint 
\begin{equation} \label{eq:comp_pow_cons}
	\tilde{\bf u}^H_k\tilde{\bf u}_k \le \frac{P_0}{ N_u P_g}
\end{equation}
Then, the per-user power constraint for the $I$ and $Q$ branch precoded sample streams can be written for $\theta \in {I,Q}$ as
\begin{equation} \label{eq:final_pow_cons}
	({\bf u}^{\theta}_k)^T({\bf u}^{\theta}_k)\le \frac{P_0}{ 2 N_u P_g}
\end{equation}
\subsection{Transmission Bandwidth Constraint} \label{app:bandwidth}

The transmitted signal at an individual TX antenna is given by \eqref{eq:TX_signal_l} where the fundamental bandwidth constraint is to limit the spectral content to $W_{TX}$ where \textcolor{black}{$1/T_{sym}< W_{TX}=\frac{1+\epsilon_{TX}}{T_{sym}} < \text{log}_2(R_{in})/T_{sym}$} based on \eqref{eq:wtx_range} where \textcolor{black}{$0 \leq \epsilon_{TX}<1$}. Given the Fourier transform of \eqref{eq:TX_signal_l}, expressed as 
\begin{align} \label{eq:FT_signal_l}
	{\phi}_{TX,l}(f)&=\int_{-\infty}^{\infty} \tilde{s}_{TX,l}(t) e^{-j2\pi ft}dt \nonumber \\
&=\int_{-\infty}^{\infty}\sum_{i=0}^{N_q-1} \tilde{s}_l(i)g_{TX}(t-iT_{u})e^{-j2\pi ft}dt \nonumber \\
&=\sum_{i=0}^{N_q-1} \tilde{s}_l(i)\int_{-\infty}^{\infty}g_{TX}(t-iT_{u})e^{-j2\pi ft}dt \nonumber \\
&=\phi_l(f)G_{TX}(f)
\end{align}
where $G_{TX}(f)$ is the Fourier transform of the transmit filter and $\phi_l(f)=\sum_{i=0}^{N_q-1}\tilde{s}_l(i)e^{-j2\pi fT_ui}$, the bandwidth constraint implies that
\begin{equation}
{\phi}_{TX,l}(f)=0, \quad \mid f \mid > W_{TX}/2.
\end{equation}

\textcolor{black}{As discussed in Appendix A, the transmit filter is assumed to be band-limited to $W_{TX,g}=\frac{1+\epsilon_{TX,g}}{T_u}=\frac{(1+\epsilon_{TX,g})\mu_{TX}}{T_{sym}}$ where $0\le \epsilon_{TX,g} \le 1$ is the roll-off factor, i.e., $G_{TX}(f)=0$, $\mid f \mid > \frac{W_{TX,g}}{2}$. When $W_{TX,g}>W_{TX}$, the TX filter does only part of the band-limitation task, while the rest must be enforced through $\phi_l(f)$, i.e., the spectral characteristics of the precoded samples.} This covers specifically the frequencies $W_{TX}/2 < |f| < W_{TX,g}/2$.

Next, we redefine the discrete time Fourier transform (DTFT) of the precoded sequence $\tilde{s}_l(i)$ as $\phi_l(f')=\sum_{i=0}^{N_q-1}\tilde{s}_l(i)e^{-j2\pi f'i}$ where $f'=fT_u$. Then, the band-limitation task to be achieved through the precoding can be expressed, in terms of the normalized frequencies, as the following constraint
\begin{equation}\label{eq:DTFT_cons}
	{\phi}_{l}(f')=0, \quad a_0 < \mid f' \mid \le b_0
\end{equation}
where \textcolor{black}{$a_0=\frac{W_{TX}T_u}{2}=\frac{1+\epsilon_{TX}}{2\mu_{TX}}$} and \textcolor{black}{$b_0=\frac{W_{TX,g}T_u}{2}=\frac{1+\epsilon_{TX,g}}{2}$}.

In \eqref{eq:DTFT_cons}, the constraints are expressed based on the DTFT of the precoded vector $\tilde{\bf s}_l=[\tilde{s}_l(0), \cdots, \tilde{s}_l(N_q-1)]^T$. Note that for each antenna branch $l$, $\tilde{\bf s}_l$ is a linear transformation of quantization precoded samples as expressed by \eqref{eq:S_mat}. Hence, the constraint in \eqref{eq:DTFT_cons} holds for each $l$, if the DTFT of each quantization precoded output stream, denoted by $\psi^{\theta}_k(f')=\sum_{i=0}^{N_q-1}u^{\theta}_k(i)e^{-j2\pi f'i}$, satisfies 
\begin{equation} \label{eq:psi}
\psi^{\theta}_k(f')=0, \quad a_0 < \mid f' \mid \le b_0.
\end{equation}
Typically, the spectral mask and other emission constraints of the system define a tolerable level of signal power that can be emitted out-of-band, and hence the strict constraint of zero frequency response in \eqref{eq:psi} can in practice be relaxed as
\begin{equation}\label{eq:DTFT_constraint}
	\int_{-b_0}^{-a_0} \mid {\psi}^{\theta}_{k}(f') \mid^2 df' +\int^{b_0}_{a_0} \mid {\psi}^{\theta}_{k}(f') \mid^2 df' \le \alpha P_0
\end{equation}  
where $0\le \alpha \ll 1$ is a constant that determines the amount of allowed out-of-band emission power. In order to impose such frequency domain constraint into the convex optimization problem with discrete variables, we finally consider the sampled spectra with integer index variable $p$, with $0\le p \le N-1$ and $N\ge N_q$, i.e.,
\begin{equation}
\psi^{\theta}_{k,N}(p)=\psi^{\theta}_k\left(\frac{p}{N}\right)=\sum_{i=0}^{N_q-1}{u}^{\theta}_k(i)e^{-j \frac{2\pi p i}{N}}.
\end{equation}   
Note that $\psi^{\theta}_{k,N}(p)$ is directly the $N$-point discrete Fourier transform (DFT) of the sequence ${u}^{\theta}_k(i)$ and the transformation can be given in the standard vector-matrix form as
\begin{equation}
	\boldsymbol{\psi}^{\theta}_{k,N}={\bf F}{\bf u}^{\theta}_k
\end{equation} 
where $\boldsymbol{\psi}^{\theta}_{k,N}=[\psi^{\theta}_{k,N}(0), \cdots, \psi^{\theta}_{k,N}(N-1)]^T$ and ${\bf F}$ is the DFT matrix with $(n,m)$'th element being $e^{-j\frac{2\pi nm}{N}}$. The conditions in \eqref{eq:DTFT_constraint} can now be adapted accordingly, which yields the bandwidth constraint used in \eqref{eq:convex_problem} as
\textcolor{black}{
\begin{align} \label{eq:comp_bandwidth_cons}
({\bf V}\boldsymbol{\psi}^{\theta}_{k,N})^H({\bf V}\boldsymbol{\psi}^{\theta}_{k,N})&=\sum_{p=p_1}^{N-1-p_1} \mid \psi^{\theta}_{k,N}(p) \mid^2  \nonumber \\
& \le \frac{\alpha (N-1)P_0}{1+\epsilon_{TX,g}}=\alpha P'_0
\end{align}
where ${\bf V}=\text{diag}({\bf v})$, ${\bf v}=[{\bf 0}_{1\times p_1}, {\bf 1}_{1\times N-2p_1}, {\bf 1}_{0\times p_1}]^T$, $P'_0=\frac{ (N-1)P_0}{1+\epsilon_{TX,g}}$ and $p_1=\lceil{p_0}\rceil$ with $p_0=\frac{a_0(N-1)}{2b_0}=\frac{1+\epsilon_{TX}}{1+\epsilon_{TX,g}}\frac{N-1}{2\mu_{TX}}$. The last line of \eqref{eq:comp_bandwidth_cons} is obtained based on the approximation $\int^{-a_0}_{-b_0}  |{\psi}^{\theta}_{k}(f)|^2 df+\int^{b_0}_{a_0} | {\psi}^{\theta}_{k}(f)|^2 df \approx \sum_{p=p_1}^{N-1-p_1} |\psi^{\theta}_{k,N}(p) |^2 \Delta_{p}$ where $\Delta_{p}=\frac{2b_0}{N-1}=\frac{1+\epsilon_{TX,g}}{N-1}$. The approximation accuracy improves with the number of used DFT points $N$.}

\begin{IEEEbiography}
	[{\includegraphics[width=1in,height =1.25in, clip,keepaspectratio]{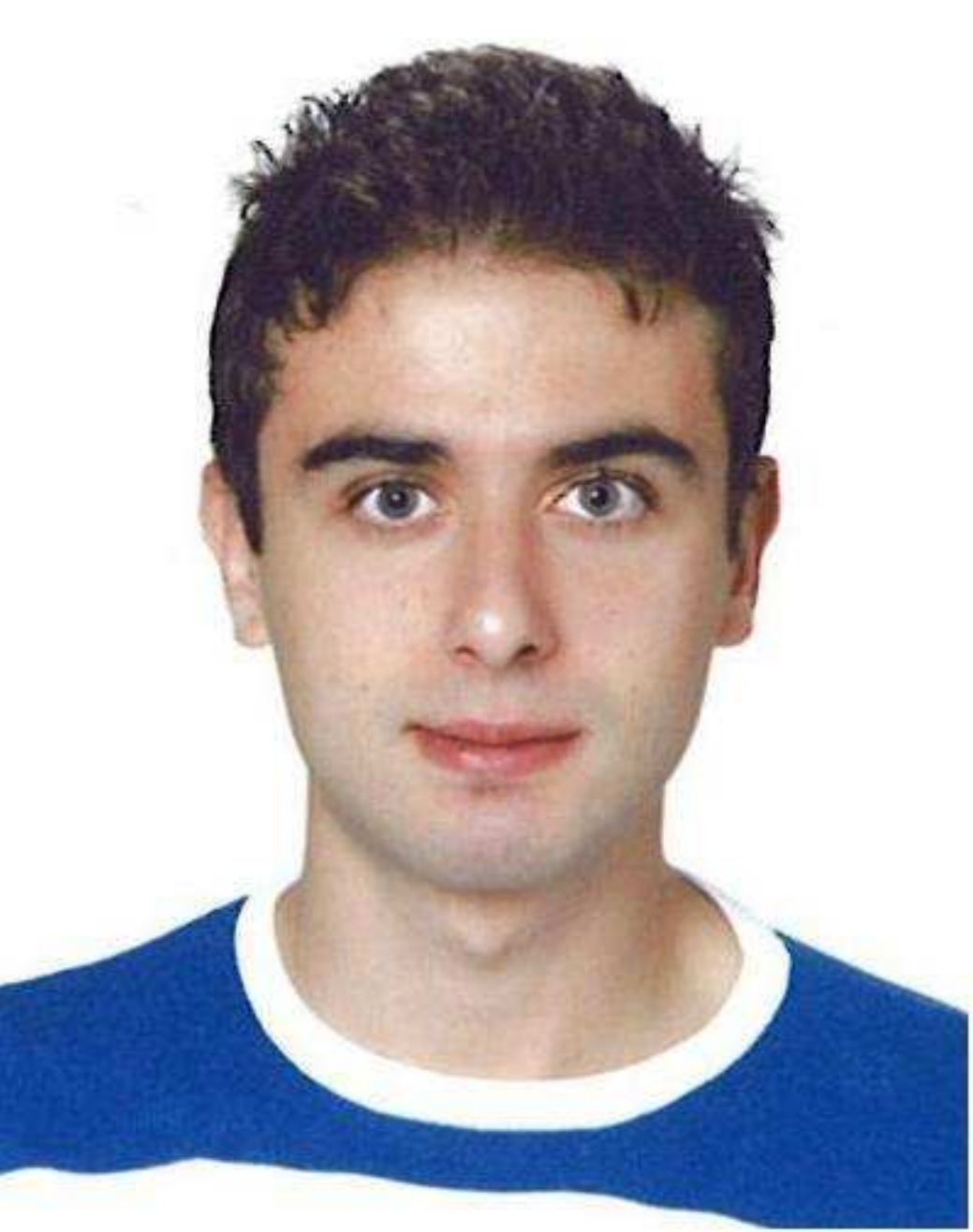}}]
	{Ahmet Gokceoglu} is a postdoctoral researcher at  Tampere University of Technology, Finland. He received the MSc and PhD Degrees from the Department of Electronics and Communications Engineering, Tampere University of Technology. His research interests are information theory, modelling and performance analysis of wireless radio communication systems, advanced transmitter and receiver signal processing, waveform optimization and 5G. 
\end{IEEEbiography}
\begin{IEEEbiography}
	[{\includegraphics[width=1in,height =1.25in, clip,keepaspectratio]{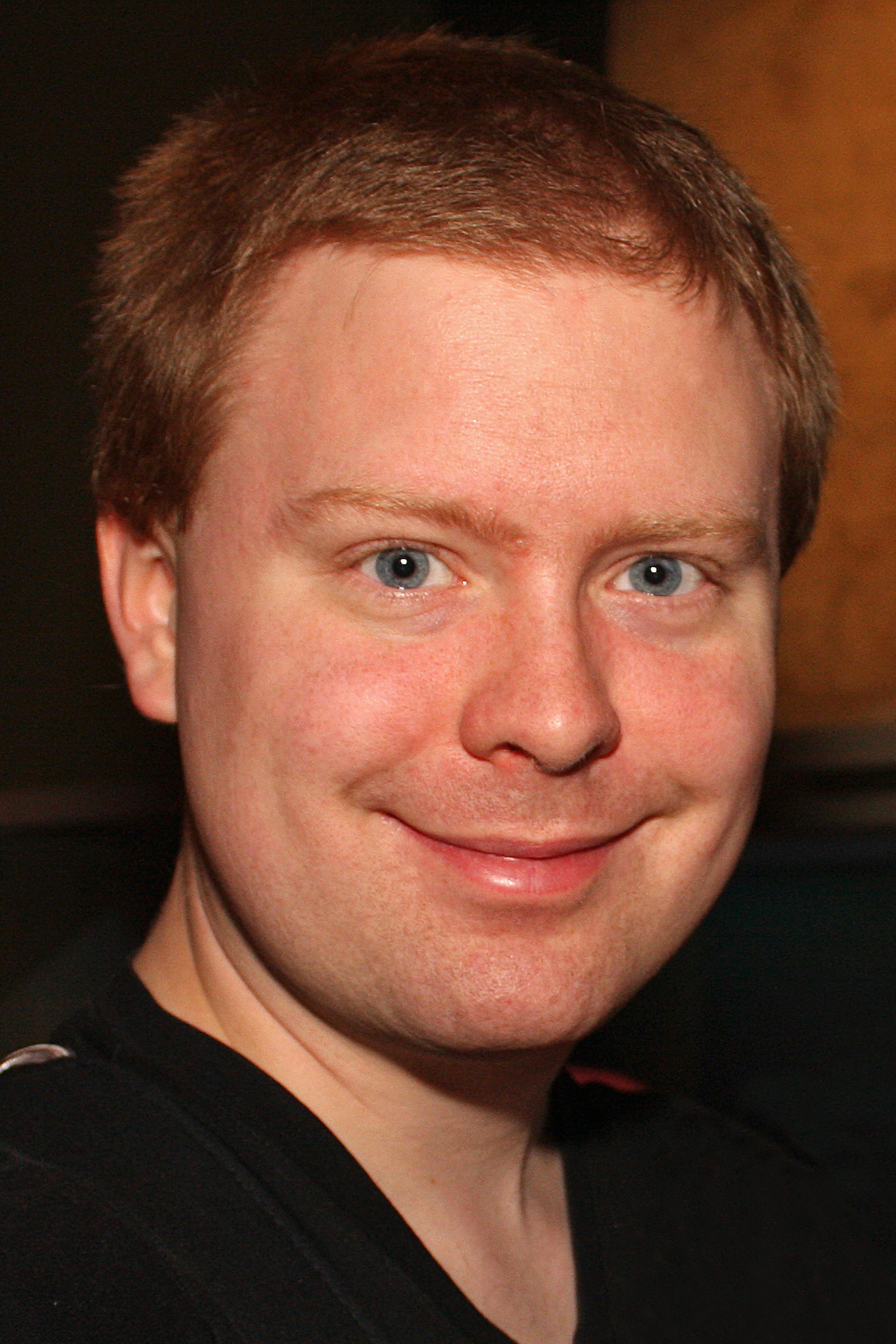}}]
	{Emil Bj\"ornson} is an Associate Professor at Link\"oping University, Sweden. He has 10-year experience of MIMO research and has filed a handful of related patent applications. He received the 2014 Outstanding Young Researcher Award from IEEE ComSoc EMEA, the 2015 Ingvar Carlsson Award, and the 2016 Best PhD Award from EURASIP, and five best paper awards. He currently serves as Associate Editor for IEEE Transactions on Green Communications and Networking. 
\end{IEEEbiography}
\begin{IEEEbiography}
	[{\includegraphics[width=1in,height =1.25in, clip,keepaspectratio]{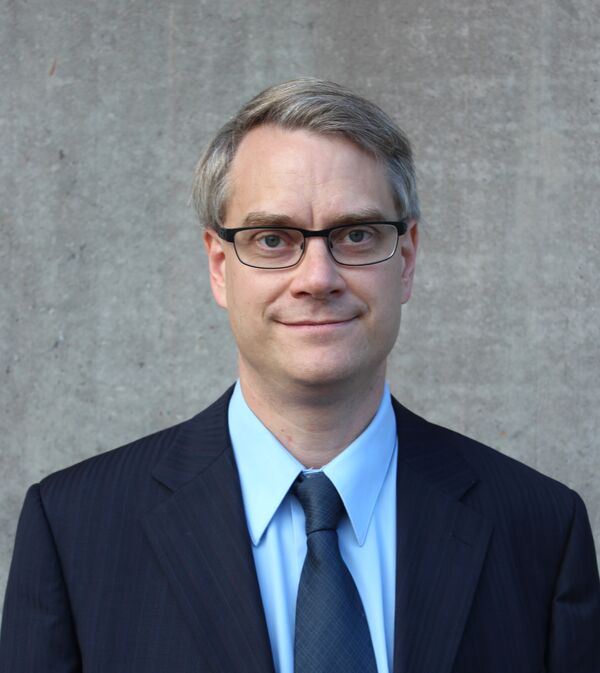}}]
	{Erik G. Larsson} is Professor at Link\"oping University, Sweden, and a Fellow of the IEEE.  His main professional interests are within signal processing, communication theory, applied information theory, wireless systems and 5G.  Recent service includes chairing of the IEEE Signal Processing Society SPCOM technical committee (2015--2016), chairing of the steering committee for the \emph{IEEE Wireless Communications Letters} (2014--2015), and organization of the Asilomar SSC conference (general chair 2015, technical chair 2012). He received the \emph{IEEE Signal Processing Magazine} Best
	Column Award twice, in 2012 and 2014, and the IEEE ComSoc Stephen O. Rice Prize in Communications Theory in 2015. 
\end{IEEEbiography}
\begin{IEEEbiography}
	[{\includegraphics[width=1in,height =1.25in, clip,keepaspectratio]{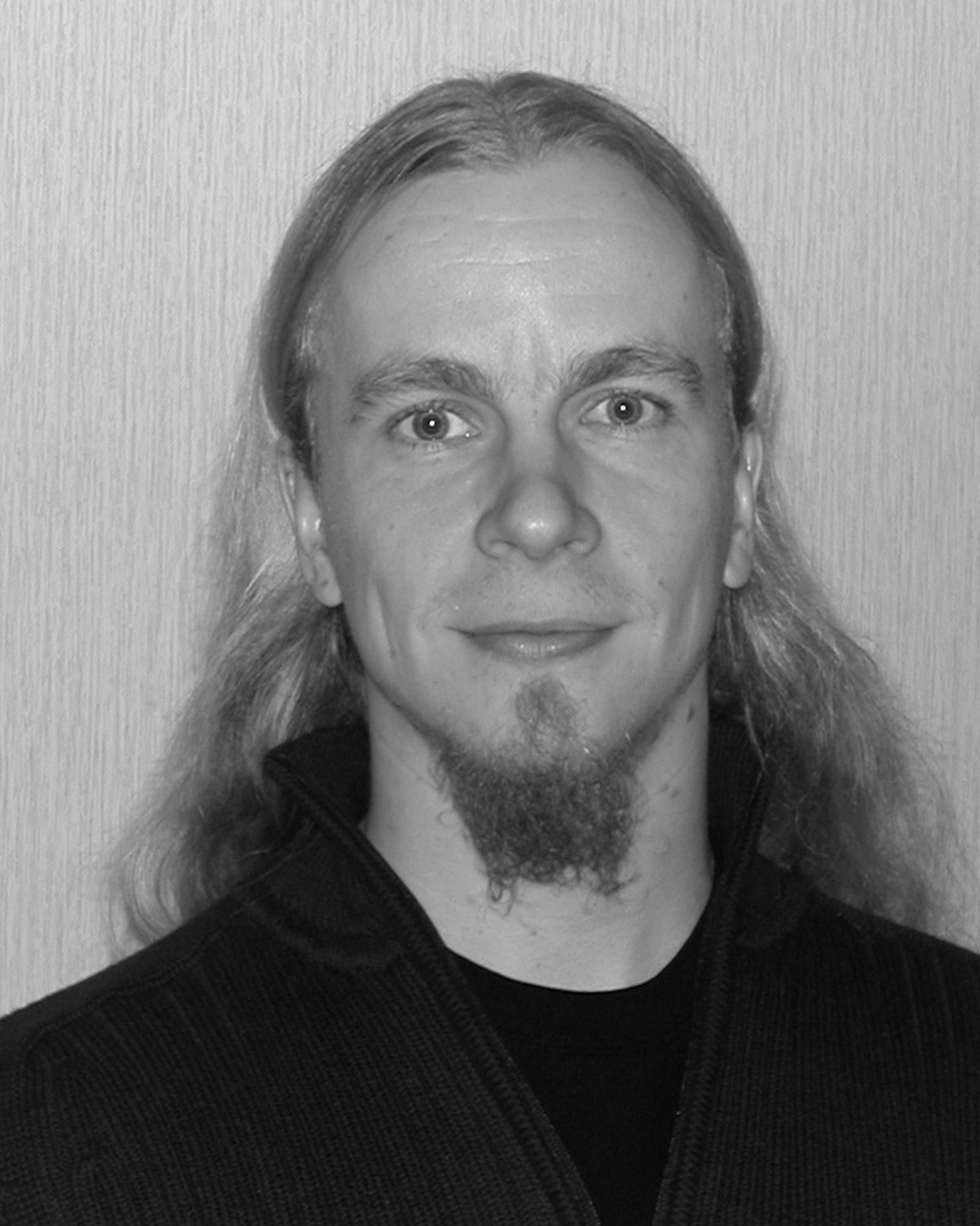}}]
	{Mikko Valkama} received Ph.D. degree (with honors) in electrical engineering (EE) from Tampere University of Technology (TUT), Finland, in 2001. Currently, he is a Full Professor and Department Vice-Head at the Department of Electronics and Communications Engineering at TUT, Finland. His general research interests include radio communications systems, networks and signal processing, with particular focus on 5G radio systems and mobile networks. 
\end{IEEEbiography}

\end{document}